# Frustration In Physiology And Molecular Medicine

R. Gonzalo Parra[1], Elizabeth A. Komives[2], Peter G. Wolynes[3*], Diego U. Ferreiro[4,5*]

1. Life Sciences Department, Barcelona Supercomputing Center, Barcelona, Spain.
2. University of California, San Diego, USA.
3 Center for Theoretical Biological Physics, Rice University, Houston, TX 77005.
4 Protein Physiology Lab, Departamento de Química Biológica, Facultad de Ciencias Exactas y Naturales, Universidad de Buenos Aires, Buenos Aires C1428EGA, Argentina.
5. Instituto de Química Biológica de la Facultad de Ciencias Exactas y Naturales, Consejo Nacional de Investigaciones Científicas y Técnicas - Universidad de Buenos Aires, Buenos Aires C1428EGA, Argentina

* Corresponding authors: pwolynes@rice.edu, ferreiro@qb.fcen.uba.ar

## ABSTRACT

Molecules provide the ultimate language in terms of which physiology and pathology must be understood. Myriads of proteins participate in elaborate networks of interactions and perform chemical activities coordinating the life of cells. To perform these often amazing tasks, proteins must move and we must think of them as dynamic ensembles of three dimensional structures formed first by folding the polypeptide chains so as to minimize the conflicts between the interactions of their constituent amino acids. It is apparent however that, even when completely folded, not all conflicting interactions have been resolved so the structure remains 'locally frustrated'. Over the last decades it has become clearer that this local frustration is not just a random accident but plays an essential part of the inner workings of protein molecules. We will review here the physical origins of the frustration concept and review evidence that local frustration is important for protein physiology, protein-protein recognition, catalysis and allostery. Also, we highlight examples showing how alterations in the local frustration patterns can be linked to distinct pathologies. Finally we explore the extensions of the impact of frustration in higher order levels of organization of systems including gene regulatory networks and the neural networks of the brain.

**1. INTRODUCTION**

In everyday life, the word "frustration" refers to the problems we humans encounter when trying to achieve multiple goals, sometimes achieving optimality for one goal rather than another. We know non living things do not really have "goals" so it comes as a surprise that the anthropomorphic term "frustration" entered the vocabulary of solid state physicists in the 1970's (Vannimenus and Toulouse 1977). By that time, theorists had achieved an understanding of how some non living systems spontaneously order at low temperatures, for example: most liquids easily crystallize and become rigid upon cooling; ferromagnetic materials orient their spins along a single direction below the Curie point. This tendency to form a microscopic repetitive pattern of order, however, sometimes fails to occur. Examples of this failure to form ordered patterns is exemplified by the fact that complex solutions often do not freeze into a beautiful crystal but settle into a still malleable glassy state. This failure to achieve optimality was a puzzle to theorists.

When looked at in molecular terms, at least after the fact, one can see why a non living system has taken on the ordered state. The ordered state is not only globally a low energy configuration but each molecular piece within it fits together exceedingly well even locally. For example, in a diamond crystal every carbon atom is beautifully tetrahedrally bonded. For more complicated substances everything resembles a jigsaw puzzle of an Escher print. The parts can readily be seen to fit together properly- but this may be clear perhaps only after you have finished the assembly, and every oddly shaped piece has found its place. In contrast, at the molecular level, a sample of a material that has formed a glass structurally makes no particular sense, and indeed, it becomes clear that many different arrangements of the molecules would turn out to be nearly equally satisfactory or unsatisfactory in energetic terms. Some interactions in a glass appear to be locally quite stable but in making these parts correct ineluctably other local interactions appear to be suboptimal and could be replaced by others. It may be said the tendency to order and satisfy the local bonding rules has been "frustrated". To quantify this inability to satisfy all the bonding rules, we can imagine searching for "frustration" in a solid state system by first examining local structures then evaluating the various contributions to the binding energies made by these local structures and finally gathering the statistics for these energies. Which parts stand out as being optimal, while poorly stabilized? In a system like a glass or a liquid a wide range of local energies of interaction will be found but for a crystal one would instead find only a narrow range of local patterns due to the periodicity and therefore also a narrow distribution of local energies. In addition, if we had sampled structures from both the crystal and the disordered states, the crystalline structure would stand out by being much lower in energy than the widely distributed energies of the disordered structures, otherwise entropy would favor the disordered glassy states.

 This physical conception of frustration can be taken over to molecular biology, but with an additional twist. While liquids have no "purpose" from the viewpoint of evolution, biomolecules actually do have a purpose: to enable the biomolecules to contribute to the survival of the organism in which they are found. So it is plausible that we could speak of frustration both as the physicists do and in the popular sense in the biomolecular context. In the last decades it has proved possible to connect "frustration" in the physicist's quantitative sense using the statistical character of the energy landscape of biomolecules, with the evolutionary requirements that biomolecules have to both fold and function in specific ways in living organisms (Ferreiro, Komives, and Wolynes 2014, Ferreiro, Komives, and Wolynes 2018). The main purpose of this review is to survey how the quantification of frustration using ideas of energy landscape theory (Bryngelson and Wolynes 1987) gives insight into the mechanisms of many biological functions by pinpointing the crucial parts of the biomolecules that allow them both to self organize into controlled structures and at the same time remain malleable so as to respond to specific signals and carry out catalysis and recognition. Medicine often deals with places where evolution has partially failed. In the context of molecular medicine we will therefore particularly highlight in this review cases where disease-causing mutations modify the frustration patterns of biomolecules in such a way as to lead to malfunction and pathology.

While this review will primarily focus on frustration in individual biomolecules and their assemblies, it has also been recently appreciated that the frustration concept may give insight into how many biomolecules cooperate with each other in regulatory networks within cells (Sasai and Wolynes 2003) and in neural networks. Artificial neural networks which are the mathematical heart of machine learning software, in fact, provide one of the clearest examples of how frustration controls the balance between function and malfunction in a complex system with a purpose. As an introduction to the concept of frustration and energy landscape we therefore will start by exploring the Hopfield model of an associative memory neural network, in the next section.

## 2. FRUSTRATION IN NEURAL NETWORKS AND THE MAGNETIC ANALOGY

The most basic operations require a brain to store vast quantities of information but then to be able to quickly access this information both in a specific detailed way (memorization) and to recognize when a similar situation occurs which should lead to the same outcome (generalization). "Free association" has been a tool of the psychiatrist for more than a century (Jung 1910). Seventy-Five years ago, a mechanistic understanding of how a simple associative memory neural network could be constructed was put forth by Hebb (Hebb 1949). Such networks could not only memorize but also generalize so as to respond appropriately to similar situations. Decades later, Hopfield showed how to connect Hebb's mechanism of memory formation and recall to the statistical mechanics of magnets and spin glasses (Hopfield 1982).

In the hyper oversimplified picture usually adopted by theorists, Hebbian neurons are viewed as having two distinct states: active or inactive corresponding to the neuron firing rapidly or slowly. These neurons communicate between each other through synapses, so that electrical and chemical signals from the firing of one neuron increase or decrease the rate of firing of its partners. There are two types of synapses, excitatory or inhibitory, in these simple models, often corresponding to chemically distinct neurotransmitters. In pairs connected by excitatory synapses, the rates of the firing of the neurons will positively correlate with each other, but for inhibitory synaptic connections, the firing rate of the two communicating neurons will anti-correlate, turning on one neuron turns off the other neuron.

It is easy to see that if all synaptic connections in a network were excitatory, the natural state of the brain as a gestalt would either be all the neurons being on or all neurons being off. it is less easy to see what would happen if a significant fraction or even all of the synaptic connections were inhibitory. Would the network freeze into a single confused but static pattern of firing, or would one get a turbulent, incoherent ever-changing pattern as in epilepsy (Wyllie et al. 2012)? The outcomes clearly depend on the detailed patterns of synaptic connection between the neurons. Hebb suggested that an exceedingly simple rule would allow the network to be configured so as to reliably recall any input pattern of firings once prompted by an input: if a consistent set of inputs to the memory network caused a set of neurons to become active, strengthening their connection between the pairs which are correlated in the input signal would make this firing pattern much easier to achieve. Of course, a brain receives many distinct and time varying sensory signals. If neurons and their synapses were continuously updated by Hebb's rule, it is clear that the network would be able to be conditioned, much as Pavlov's dogs were trained to salivate when a bell rings (Nobelstiftelsen 1967).

Hopfield translated Hebb's rule into the language of magnets and solid state physics (Hopfield 1982). Magnets often can be thought of as crystals where electron spins live on each site. If isolated, each spin could either be pointing up or down, just as a neuron can be stylized as actively firing or not firing. Spins may couple to each other in energetic terms through the laws of chemical bonding. The pattern of the coupling between the spins depends on the local chemistry which may energetically favor the spins being parallel to each other- ultimately giving a ferromagnet as the lowest energy state or (as is more familiar to organic chemists in bonding where free radicals are reactive) align themselves to be antiparallel so that the spins cancel macroscopically. The energy of such a system of N spins each taking on the value + (or -) labeled for i=1 to i=N would be $E = -\sum_{i \neq j}^{i=n} J_{ij} S_i S_j$ . So $J_{ij} < 0$ favors parallel spins, $J_{ij} > 0$ favors cancelling spins. If all $J_{ij} > 0$, we have a ferromagnet. If all $J_{ij} < 0$ we have an antiferromagnet. Hopfield realized that the Hebb rule allows one to make a magnet that can memorize any given single pattern no matter how diverse is its spin orientation by choosing $S_i$ to give $J_{ij} \equiv -J_0 S_i^1 S_j^1$ any complicated pattern $S_i^1$ can then be memorized by the system once the $J_{ij}$ patterns have been set. The thermal motions of the spins will settle into a low energy state with the spins being ordered up and down just as they are in the "memorized" training example. More than one pattern can be memorized by such a system of spins. By choosing the $J_{ij}$ values as the average of the desired training memories $J_{ij} = \frac{1}{M} J_0 \sum_{\alpha=1}^{M} S_i^\alpha S_j^\alpha$ the spin dynamics would favor all of these input memorized patterns of spins. While two such patterns might even be inconsistent with each other, putting in a very weak additional signal will result in the system settling into one of the learned patterns just as in the same way weakly jingling the dog's bell in the Pavlov experiment causes the dog to salivate. The spins will take on one of the memorized patterns at low temperature. Nevertheless, the shared memories may be in conflict, so there is diversity in the recall.

There is a limit to the ability of the network to reliably recall its training: If the number of memorized patterns M is too large compared to the number of neurons N, recall breaks down. At first when the network is only a little above its capacity such an overtrained magnet "hallucinates": the system does not recall any specific input pattern but wanders into some combination of learned states (much like a dream where one of your friends turns into a bat! (Freud and Brill 1915). At still larger values of M, the magnet rather than hallucinating the spin system freezes into a random pattern, one of an exponentially large number of states none of which resembles any of the input patterns.

This frozen state of an overtrained network is called a "spin glass" because the dynamics of cooling such a magnet corresponding to an overtrained network resembles the way molecules haphazardly freeze into amorphous structures found in the solid glasses we encounter in everyday life. In such glasses many local energetic interactions are stabilizing but others are not. In the Hopfield magnetic analogy, we use the term "frustration" to refer to the energetic inconsistency of different patterns of spins to be stable memories. The specific behavior of the system under recall then depends on the relative energies of the patterns to be recalled in comparison to the glassy random patterns which are much more numerous and may by chance compete in a thermodynamic process with the proper patterns to be learned.

Proteins, in many ways, resemble the Hopfield memory magnet. Under some conditions, high temperature or high denaturant concentration, proteins in their thermal motion scramble across a rough energy landscape and are said to be unfolded. At low temperatures, they fold into a much more organized, "folded" state with many stabilizing interactions that work together consistently. This folded state is still a large ensemble of structures, resembling a hazy memory recall. Some of this "haziness" is useful because the ambiguity gives mobility and allows function: if hemoglobin were a completely frozen structure like the one revealed by X-ray crystallography, oxygen couldn't get in to bind!

Sometimes, a protein has a few dissimilar but highly related structural ensembles that might have many common features, owing to their similarities, each form of the protein can do the same work but with differing efficiencies. This is often the route to "allostery" which is so important to molecular biology. Allostery thus resembles the hallucinating phase of Hopfield associative memory neural networks.

The analogies between solid-state systems, neural networks and proteins have been explored and exploited in many ways. The realization that the number of basic patterns of global folds is quite modest implies that as a computational task structure prediction from sequence can be considered an association problem: one needs to associate a given sequence to an appropriate three dimensional structure: similar sequences fold to similar structures. This way of looking at structure prediction has inspired the use of neural networks to recognize tertiary structures (Friedrichs and Wolynes 1989), leading eventually, with the growth of sequence data, to today's powerful neural networks for structure prediction (Jumper et al. 2021), which are often called AI or "Artificial Intelligence." The analogy between solid state systems and proteins also has given insight into many mysterious but common aspects of protein folding mechanisms and the unusual kinetics of folding and unfolding that can be quantitatively understood and predicted with the mathematical tools of energy landscape theory (Peter G. Wolynes 2005; Oliveberg and Wolynes 2005).

One of the ways proteins seem to differ from the biological neural networks of the brain is in the way the interactions have been trained in nature. For proteins, it is clear that folding and proper function arise from natural selection by many generations of organisms reproducing and then selectively dying, weeding out those expressing unfoldable molecules. Unlike the Brain there is very little active "learning" that proteins do in real-time, at least individually. Perhaps, cytoskeleton networks of proteins do display some form of active structural learning when dendritic spines reconfigure after synaptic inputs (Gu et al. 2020; Liman et al. 2020). In contrast, a proposed role of evolutionary memory and inborn behaviors (archetypes) in psychology has certainly been considered controversial ever since it was put forward by Jung. Because protein dynamics was learned by evolution we will see that quantitatively analyzing the local energy landscapes of proteins very usefully teams up with evolutionary sequence analysis (Peter G. Wolynes 2015; Morcos et al. 2014).

## 3. LOCAL FRUSTRATION IN PROTEINS

Admittedly natural protein molecules are beautiful. The last 60 years of experimental resolution of protein structures and the now millions of predicted structures that we compile and visually explore, amaze us in the variety and subtleness of the structural forms that natural proteins adopt. Some of us are still impressed by the fact that such a variety of structures can be made with aperiodic repetitions of just 20 basic building blocks. These forms are unlike the mathematically perfect forms of inorganic crystals but we can still detect recurrent patterns that pack in non symmetrical manners that somehow calls our aesthetic attention. The process by which the amino acid chains adopt the stable structural forms is appropriately called 'Protein Folding'. How does this process occur? The major advance in our understanding of proteins is the Energy Landscape Theory of protein folding, which postulates that the energy landscape navigated by natural proteins has the form of a rough funnel (Onuchic, Luthey-Schulten, and Wolynes 1997). Funnels are possible for systems that follow the Principle of Minimal Frustration, for which there is a fundamental correlation between similarity in conformation and energetic distribution (Bryngelson and Wolynes 1987). To gain some insights about these processes, let's start with a simple example: an homo-polypeptide, say a stretch of 100 poly-alanine. How do you imagine that this polypeptide would fold? What is the relation between structure and energy? Lets imagine we compact the polypeptide into a coil and we measure the energy as the sum of the interactions between the amino-acids. In this case, since they are all alanine, there would be an average energy given by the sum of the pairwise interactions in the compact state. Now let's imagine we compact the chain into another structure. And another, and yet another. No matter what structure we fold the chain into, most of the alternative structural forms will have similar energies, and no structural discrimination is possible. The landscape associated with a homo-poly-peptide is very rough. If we start an experiment at high temperature and let the polymer explore the conformation space, it will explore a large variety of structures. As we cool the system, every copy of the molecule can settle in different local minima and at some point there will be no kinetic energy to be able to overcome the barriers that separate them. The system will settle into a glass, populating many different structural forms of similar energies. This is not how a natural protein behaves, and indeed natural protein molecules are not regular homo-polymers. Now let's imagine the opposite case, a polypeptide made by attaching together 100 amino acids thrown at random from the pool of the genetically coded 20. How would this system behave? Following the same scheme we used in the previous case, we could ask what is the energetic expectation of variously collapsed structures. Can you see that the expected energy landscape is also rough? Collapsing a random heteropolymer into different forms will nearly always lead to conflicts in the interactions between amino acids. Burying for example hydrophilic residues in a globule of hydrophobic ones to which they are linked. The system is very frustrated. Within a random polypeptide, there is no possibility of encoding a robust structural form. So, natural protein molecules are neither homopolymers nor just chance sequences of amino-acids. There must be energetic correlations such that the system is guided to the energy minimum that is reproducibly observed. Such correlations must account for the folding codes that lead to a correspondence between sequences and structures.

The fundamental distinction that natural proteins are the result of evolved sequences led to the proposal of a rigorous mathematical framework to infer the energetics of protein folding, a license to learn physics via bioinformatics. If natural proteins are minimally frustrated heteropolymers, the energies that stabilize the folds must be sufficiently higher than the energies found in random compact states. By adjusting the energetic fields such that the discrimination between native and random compact states is maximized, reliable energy functions can be constructed based on the theory of neural networks (Friedrichs and Wolynes 1989). Over the decades, these ideas developed into a consistent associative memory, water mediated, structure and energy model (AWSEM) as a coarse-grained protein force field (Davtyan et al. 2012). This energy function has proved capable of performing de novo protein structure prediction. Moreover, since the force-field contains physically motivated terms, it can be used for kinetics and dynamics studies, and to evaluate how mutations or structural changes affect the protein energetics.

Having a reliable way of estimating the energy for a given sequence folded into some structure, we can ask how does the energy change when we make changes to the sequence or the structure (Ferreiro et al. 2007). For this we can take a high resolution structure of a natural protein and sequentially change the pairs of amino acids that are in contact. Locally this resembles what would happen if a random collapse had occurred. For contact number n, we make all the possible 399 ($20^2$-1) mutations and for each one we measure the energy of the molecule. We can then compare the native energy to the distribution of such decoy structures and ask: does the native pair substantially contribute to stabilizing the molecule, is it average or is it destabilizing? We do this computationally for all the amino acid pairs that are in spatially contact proximity. If the interaction is sufficiently more favorable than would occur by chance, we call this site 'minimally frustrated'. If the native energy is near the mean of decoy energies we call this interaction 'neutral' and if the interaction is unfavorable we call it 'highly frustrated'. With these data we can then visualize the optimal and suboptimal parts of the structure by coloring the interactions in the structure and observing the patterns. We call this heuristic local 'mutational frustration' calculation. An alternative way of measuring the local frustration imagines that the residues are not only changed in their identity but also displaced in structure. The structure then is computationally reshuffled to generate the decoy distributions and the native energy of the interacting residues is compared to it. We call the results from this method local 'configurational frustration'.

At first a dataset of non-redundant single domain proteins was analyzed (Ferreiro et al. 2007). In Fig. 1, we show a schematic representation of the frustration results which we will now explain. In general it was observed that about 40% of the interactions fall into the minimally frustrated class. These interactions are typically shielded from solvent, constituting the core of the globules. These are more stabilizing than could be expected at random. About 50% of the interactions are described as being neutral; those where the native interactions are neither stabilizing or destabilizing with respect to the changes, and are randomly distributed along the structure. The remaining 10% of the interactions fall in the highly frustrated class. These are interactions where most of the changes would not destabilize but rather   stabilize the molecule. Obviously these do not help the folding process. These interactions typically occur in patches in the surface of the globules. Because of their instability, these highly frustrated interactions must be held in place by the rest of the structure, otherwise they would form other structures. So why would a natural, evolved protein have regions where there are unfavorable interactions between their constituting amino acids? A key to this question comes from the realization that proteins are not 'optimized' only to fold but they are the outcome of evolutionary processes in which they also must 'function'. Physics doesn't care about functions. What is the function of the moon orbiting the earth? Function is a concept genuine to Biology, where a multitude of chemical forms come together and perform some act that is related to the emergence of a phenomenon at a higher level of description. It is always difficult to disentangle what are the biological needs and necessities of a biological object and what is just a chance outcome of a noisy physical system. Which aspects of the functional forms of proteins may conflict with robust folding? Why are these frustrated interactions held over physiological and evolutionary times in protein molecules? A simple thought experiment may clarify the problem.

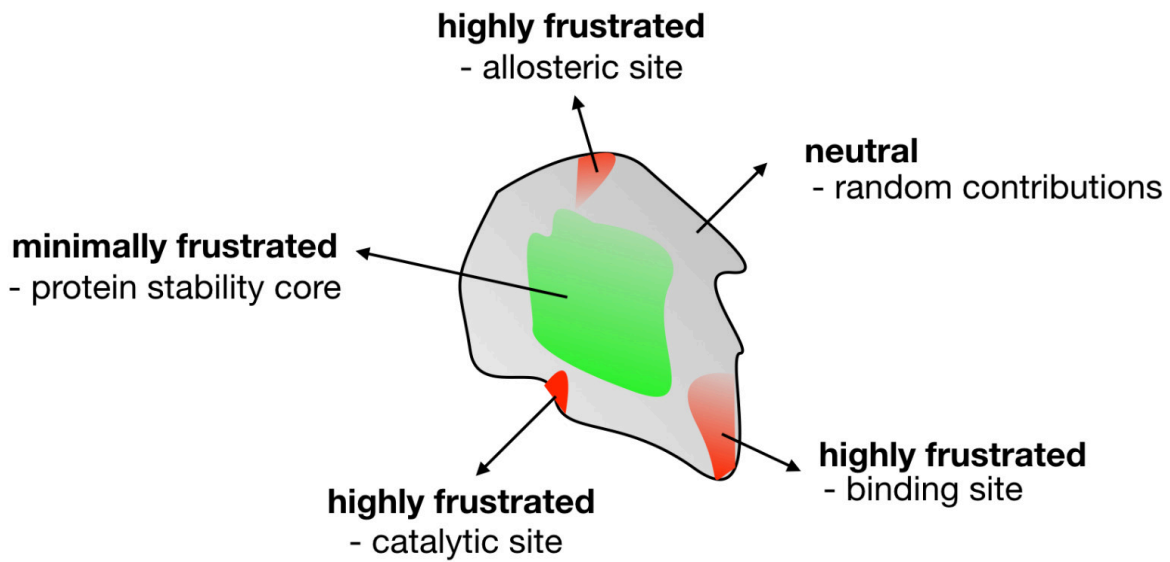

**Figure 1.** Schematic representation of the different frustration classes in a globular protein. On average, 10% of interactions are highly frustrated and are clustered, often on the surface of proteins. 40% of the interactions are minimally frustrated and typically found at the hydrophobic core. The rest of the interactions, 50%, are neutral.

## 4. PROTEIN FRUSTRATION IN PHYSIOLOGY

Proteins do not act alone. They often come together and form specific protein complexes. In general, these should be folded in order to avoid a non-specific interaction mess. Imagine that we have two well-folded globules that must interact with each other forming a protein-protein complex. To be able to interact, there must be some chemical complementarity between the objects, otherwise the interaction would not be stable enough to overcome the entropic penalty for dimer formation. This complementarity must physically come from the interactions between the amino acids that make up the surface of the globules, thus restricting the possible sequences of the polypeptide chains at the interaction region. When we observe the globules out of the complex it may happen that those positions now would appear as 'highly frustrated', as the explored changes in the monomer are agnostic to the interaction between the partners, and other sequences or structures of the partners may be needed to stabilize the fold. Indeed, when analyzed over a database of protein-protein complexes, those regions involved in specific binding were found generally to be enriched in highly frustrated patches when the partners in the complexes are artificially separated (Ferreiro et al. 2007). Moreover, when the whole complex is analyzed it is found that the overall frustration decreases, as the chemical complementarity now comes into place in the energetic analysis. The local frustration change thus guides specific protein-protein associations. A corollary of this analysis is that protein-protein surfaces cannot be too large with respect to the size of the domain, as the sequence restriction can be too demanding for the rest of the polypeptide to code for a folded conformation in isolation. In line with this is the known fact that the formation of large interacting surfaces is closely associated with folding upon binding of the interacting partners which by themselves might be deemed as 'intrinsically disordered' (Mészáros et al. 2007)

But association is not the only functional requirement that protein molecules have. Another major characteristic of proteins is that they excel at catalyzing chemical reactions. To do so, specific residues must come in close proximity with the reactants, juggling with sub-nanometer precision the atoms that must react. We can thus expect, and it has been experimentally observed, that mutations at the active sites of enzymes may stabilize the folded structures, albeit making them catalytically inactive (Shoichet et al. 1995). Analysis of a database of annotated active sites in a thousand of natural enzymes has shown that the catalytic sites themselves are often highly frustrated regardless of the protein oligomeric state, overall topology, and enzymatic class (Freiberger et al. 2019). At the same time a secondary shell of more weakly frustrated interactions usually surrounds the catalytic site itself. These might be tuned so as to facilitate multi-steps mechanisms of catalysis. There is a functional constraint that shows up in the enzymes conflicting with the necessity to code for folding.

But association and catalysis are not the only functional requirement of protein molecules. As the molecular physiologist Dennis Bray said "Many proteins in living cells appear to have as their primary function the transfer and processing of information, rather than the chemical transformation of metabolic intermediates or the building of cellular structures." Monod recognized that computations can be brought about by allostery, and maintained this phenomenon could be considered 'the second secret of life', second only to the genetic code (Monod 1971). Allostery can be defined in functional terms as a comparison of how one ligand binds in the absence, versus the presence, of a second ligand. Most importantly these ligands may be not chemically related. Take the prime example of the LacR protein. This protein is a transcriptional repressor of the lac operon in *E. Coli*. In the absence of lactose it specifically binds to a genomic region termed *lacO* operator sequence, blocking transcription. In the presence of lactose the affinity of LacR for *lacO* significantly drops, and thus transcription of the *lac* genes is relieved. The fact that lactose (a simple sugar) is related to the expression of the genes that allow the bacteria to consume it, is an outcome of the allosteric mechanism acting in LacR. There is nothing chemically intimate relating lactose to the structure of the binding site of *lacO*. In fact, many biotechnological applications have engineered this same system to regulate the expression of arbitrary genes. Thus, allosteric proteins must occupy structural forms that are malleable, and that change in response to external stimuli. One can anticipate that local frustration should play a role here. Exploring a database of known allosteric proteins, for which structures in both forms were known, it was found that the regions of the proteins that undergo structural changes are indeed enriched in highly frustrated regions (Ferreiro et al. 2011). These regions must somehow be locally unstable to be able to switch forms. However, evolutionarily engineered frustration is not the only route to allostery. The classic Wyman-Monod view of symmetry of multimeric assemblies leading to nearly energetically equal alternative states does not depend on local frustration but from degeneracy of symmetric packings (P. G. Wolynes 1996).

Overall we see that local frustration, far from being a 'bad' thing for proteins, is an essential part of their inner workings. Far from being static marble sculptures, as a naive view of the structural models suggest, proteins are molecules and as such they cannot be static entities at physiological temperatures. Proteins are 'kicking and screaming stochastic molecules' as Gregorio Weber reminds us (Weber 1992). The static pictures we get from high resolution experiments or computational modeling are just an initial step in the beginnings of our understanding of the actual protein dynamics. Natural proteins in their native state visit ensembles of structures, and the occupation of conformational substates has long been recognized as essential to their biological function (Frauenfelder, Sligar, and Wolynes 1991). Local frustration must be somehow related to the dynamics of protein molecules. A prime example of this connection has been documented in the Thrombin enzyme, which we describe below.

## 5. EXAMPLES OF FUNCTIONAL ASPECTS OF LOCAL FRUSTRATION IN PROTEIN SYSTEMS

By following the principle of minimal frustration, proteins evolve to be both thermodynamically and kinetically foldable, ensuring efficient navigation of their energy landscapes toward a stable native state. As discussed in the previous section, natural foldable proteins can tolerate a limited degree of local frustration—around 10% of highly frustrated interactions—without significantly compromising their overall stability. These frustrated regions often serve functional roles, enabling flexibility, allostery, or interactions with other biomolecules. However, a substantial increase in frustration would disrupt the folding process by introducing competing non-native interactions, creating kinetic traps, or even preventing the formation of a sufficiently stable structure altogether.

While many proteins achieve a well-defined native state on their own, others require additional mechanisms to fold properly or to perform their function. Some proteins fold only upon binding to a partner. Others rely on molecular chaperones to navigate their rugged energy landscapes, avoiding aggregation or misfolding. Disulfide bonds can sometimes further stabilize structures by reducing entropic costs, particularly in oxidative environments and at other times promote local flexibility with functional consequences. Beyond these mechanisms, more complex cases challenge the classical view of protein folding: metamorphic proteins can adopt multiple distinct folds, fuzzy interactions involve structural ensembles rather than a single conformation, and intrinsically disordered proteins (IDPs) remain unfolded or only partially structured under physiological conditions. In the following sections we will illustrate these diverse strategies that illustrate how nature balances the use of local frustration with stability and adaptability to achieve complex functions.

### 5. 1 Frustration In Nfκb-DNA Binding

NFκB is a family of transcription factor proteins that respond to cellular stress by turning on hundreds of genes involved in the immune response and in cell growth (Ferreiro and Komives 2010). By combining experiment and theory, it has been shown that NFκB signaling is controlled by a kinetic mechanism in which newly synthesized inhibitor, IκBα, irreversibly removes NFκB from DNA target sites in a process referred to as "molecular stripping" (Bergqvist et al. 2009; Potoyan, Zheng, Komives, et al. 2016). This process involves folding and rearrangement of part of IκBα including the $5^{th}$ and $6^{th}$ ankyrin repeat domains and the PEST sequence (Potoyan, Zheng, Ferreiro, et al. 2016). Once the PEST sequence was identified as a control feature of molecular stripping, it was proved possible to design a stripping deficient mutant of IκBα which could be introduced into cells. The rate of molecular stripping was thus shown to actually control the rate of nuclear export of NFκB (Dembinski et al. 2017).

NFκB dimers bind a broad range of different DNA sequences. How does this occur? Nearly half of the sites bound by NFκB in cells do not correspond to the "consensus" motif (Udalova et al. 2002) binding sequence (Siggers et al. 2011). The dimer affinity is strong enough that for most of the family members, dimerization precedes DNA binding (Siggers et al. 2011; Ramsey et al. 2019). Structures of NFκB bound to DNA all show the two DNA binding domains "hugging" the DNA and the DNA is engaged by not only the DNA-binding domains, but also the dimerization domains and the linkers that connect the DNA-binding domains to the dimerization domains (Fig. 2A). Molecular dynamics simulations show that in the absence of DNA, the DNA-binding domains adopt a full range of states with different separation distances between the two DNA-binding domains (Fig. 2B). Frustration analysis (Fig. 2C) shows that much of the DNA binding cavity is highly frustrated in the open, unbound form (W. Chen et al. 2021). Frustratometer analysis of NFκB revealed that the linkers that connect the DNA-binding domains to the dimerization domains are also highly frustrated. In addition, the surfaces of the DNA-binding domains are highly frustrated. This pattern of frustration arises largely because both the linkers and the DNA-binding domains are positively charged and would repel each other in the absence of negatively charged DNA.

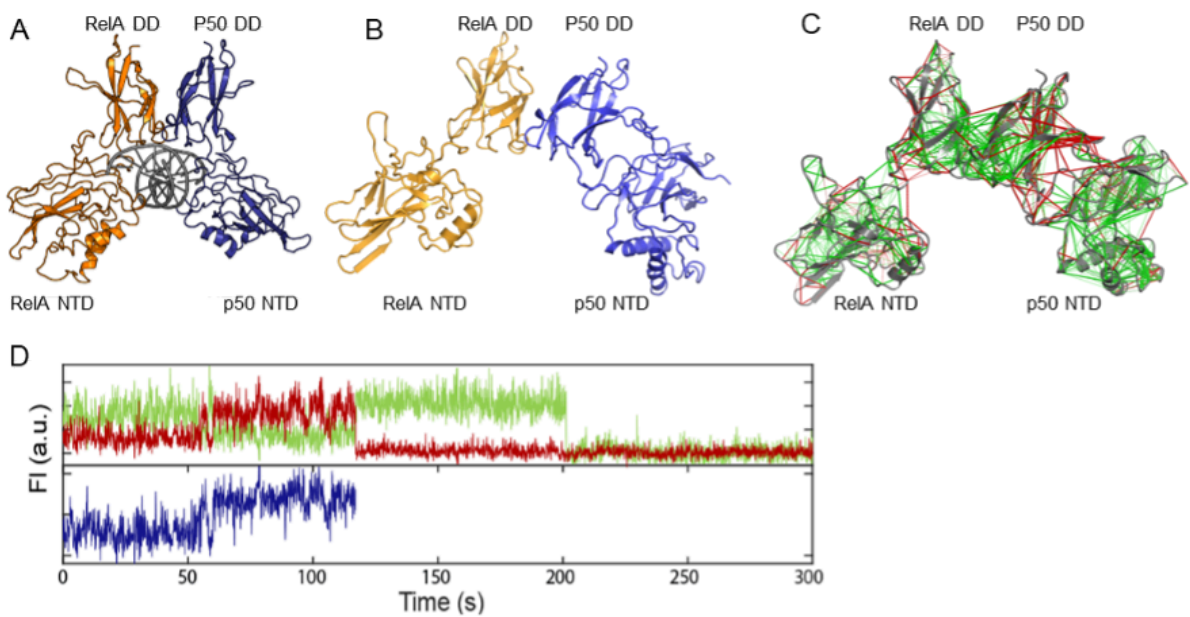

**Figure 2. A.** Structure of NFκB(RelA-p50) bound to DNA (PDB code 1LE5). **B.** NFκB(RelA-p50) after removal of DNA and 300 ns of conventional MD simulations. **C.** Frustration patterns calculated on the 1LE5 structure. **D.** An example smFRET trace from a single molecule of DNA-bound NFκB showing fluctuations between low and high FRET states are stochastic.

Single molecule FRET (smFRET) experiments show there is a stochastic interconversion between many different states from the "closed" conformation (similar to the conformation when DNA is bound) to the "open" conformation (similar to that observed in the MD simulations) on a range of timescales. Remarkably, single molecule FRET studies of κB target DNA-bound NFκB showed that even when bound to DNA, the DNA-binding domains still adopted different conformations, and often apparently only one of the DNA-binding domains was bound (Fig. 2D). The continued dynamics of the DNA-bound state is most likely due to incomplete satisfaction of the frustration within the linkers that connect the dimerization domains to the DNA-binding domains. Our results provide an explanation for the way frustration in NFκB potentiates it for IκBα-mediated 'molecular stripping'.

## 5.2 Frustration in thrombin

Thrombin is the terminal protease in the cascade of blood coagulation proteases. The structure of thrombin consists of a double β-barrel core surrounded by connecting loops and helices. Compared to chymotrypsin, thrombin has more extended loops that are thought to have arisen from insertions in the serine protease that evolved to impart greater specificity. Thrombin activity is controlled by binding at both the active site and at a distal exosite. The substrate for thrombin under pro-coagulant conditions is fibrinogen, which engages both the active site and the exosite. Under anticoagulant conditions, thrombomodulin binds at the exosite and primes the active site of thrombin to cleave protein C instead of fibrinogen. NMR relaxation experiments on the timescales of motion within thrombin revealed that its slow time-scale motions are linked to catalysis (Handley et al. 2017; Peacock et al. 2021). It is interesting that nanosecond time scale dynamics do not correlate with protein enzyme catalysis, which occurs on the millisecond to seconds timescale. Catalysis typically requires correlated loop motions that are sampled by NMR relaxation dispersion experiments. In the case of thrombin, one can compare thrombin in its apo state to thrombin bound to thrombomodulin, which promotes catalysis of protein C cleavage. It was shown that thrombomodulin completely remodeled the dynamics of thrombin marking out allosteric pathways of correlated motions in the thrombomodulin-bound form (Handley et al. 2017; Peacock et al. 2021). Analysis of the residual local frustration shows correlation with the NMR results. Using the "configurational frustration" index made it possible to discover whether regions of high frustration correlate with dynamic regions. The average residual frustration across the three lowest energy structures derived from the AMD simulations can be compared to the NMR motional order parameters. The order parameters derived from conventional MD simulations ($S^2_{ns}$) agree very well with order parameters derived from NMR relaxation experiments on thrombin, and they reveal the disorder resulting from motions in the ns time regime. While the $S^2_{ns}$ do not correspond well to the regions of high residual frustration (Fig. 3A, 3B, grey points). The order parameters derived from the residual dipolar coupling measurements combined with accelerated MD simulations ($S^2_{AMD}$) reveal the disorder resulting from motions on longer timescales proving a striking correspondence with the S2$_{AMD}$ (Figure 2B) (Fuglestad et al. 2013). These results for thrombin are consistent with an allosteric mechanism in which frustrated contacts at the allosteric site are engaged by the binding of the allosteric regulator, and then the change in dynamics is transmitted through the minimally-frustrated core of the protein to the active site (Torres-Paris et al. 2024). This same phenomenon occurs also in urokinase plasminogen activator, another serine protease involved in the blood coagulation (Torres-Paris et al. 2024).

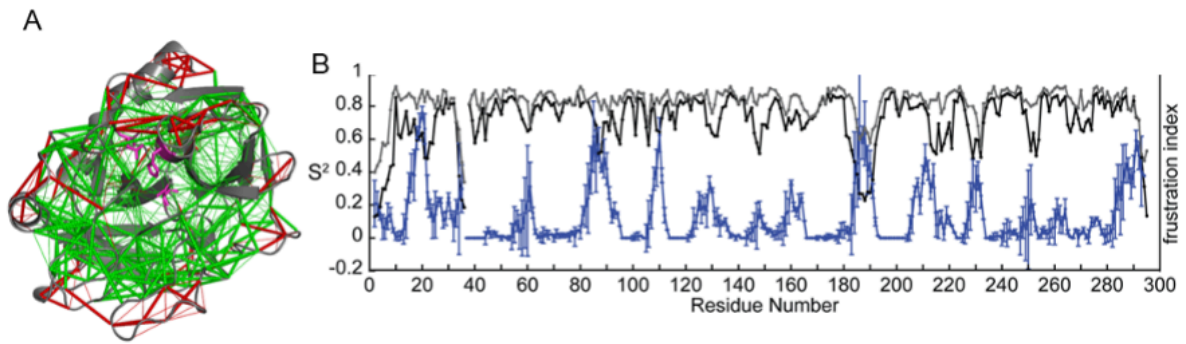

**Figure 3. A.** Frustration analysis of thrombin reveals highly frustrated surface loops. **B.** Comparison of the average fraction of highly frustrated contacts (cyan) with the S2ns (gray) and S2AMD (red) for apo-thrombin.

### 5.3 Frustration and Fuzziness in Protein Interactions

The concepts of frustration and fuzziness are intertwined in protein folding and interactions. Frustration arises from suboptimal interactions within the protein structure or complex, leading to a locally rugged energy landscape. This deviation from an ideal funnel-shaped landscape can result in diverse conformations and increased flexibility. Fuzziness, on the other hand, describes the inherent structural diversity observed in protein complexes, particularly those involving intrinsically disordered proteins (IDPs) (Fuxreiter 2018). IDPs lack a stable, single three-dimensional structure and can adopt multiple conformations upon binding to different partners. The connection between frustration and fuzziness lies in their shared origin: the ruggedness of the energy landscape. Frustration in the binding landscape can directly contribute to fuzzy binding, where proteins interact with multiple partners in a context-dependent manner (Gianni et al. 2021). This adaptability allows proteins to engage in a variety of interactions, enhancing their functional repertoire. For example, frustration at the binding interface can enable IDPs to adopt different conformations with different partners, facilitating precise and context-specific regulation (Freiberger et al. 2021).

It is crucial to recognize that not all protein interactions are fully optimized. Suboptimal interactions, often associated with frustration, do not necessarily compromise specificity. Instead, they can provide a mechanism for achieving diverse biological functions. By allowing for multiple sets of suboptimal contacts, frustration can encode specificity without requiring a single, well-defined bound conformation. This versatility is crucial for adapting to different cellular conditions and expanding the biological repertoire of proteins. Furthermore, evolutionary selection may actively favor frustration in certain cases. By maintaining a degree of frustration, proteins can retain the flexibility to interact with multiple partners or respond to changing environmental conditions. This suggests that not fully optimized structures can be advantageous for biological activity, providing a broader range of functional capabilities.

## 6. LOCAL ENERGETIC FRUSTRATION AND THE CHAPERONES - CLIENTS INTERPLAY

The level of frustration required for function could make folding unviable and thus put proteins at the risk of aggregation. Chaperones could help to avoid this problem by assisting proteins to fold when function related frustration would put proper folding at risk. Thus, it is interesting to explore the correlation between frustration and chaperones.

Indeed, many chaperone protein clients turn out to be conformationally diverse, i.e. they are not fully folded or they exhibit a certain degree of disorder during the interaction with the chaperone. The chaperone-client interactions are transient, forming a complex with multiple conformations which protects the client from aggregation and thus enhances the navigation of the energy landscape towards the folded state *in vivo* (He and Hiller 2018). This conformational diversity of clients does not require structural complementary to form the chaperone-client complex and may help to understand chaperone promiscuity and their broad clientome (He and Hiller 2018). An explanation for this promiscuity has been proposed in recent years stating that chaperones do not recognize specific sequence motifs but instead recognize and bind to local frustrated sites in their client proteins. Several holdases have been described to function following this mechanism by which they stabilize conformations that otherwise would be unstable in the absence of the chaperone (Hiller 2019).

### 6.1 Chaperone Spy

The ATP-independent periplasmic chaperone Spy (spheroplast protein y) can bind to native, folded proteins as well as to non native proteins but has a higher affinity for the non folded forms. This feature enables client proteins to fold while bound to Spy (Mitra et al. 2021). This chaperone forms a cradle-like structure through an antiparallel coiled coil interaction of two 16 kDa monomers (Quan et al. 2011). The cradle-like binding site is mostly formed by positively charged residues and allows Spy to bind periplasmic or outer-membrane proteins allowing them to fully fold. When binding its client proteins Spy does not undergo large structural alterations beyond the increased flexibility at the linker loops which in turn facilitates interactions with different substrate conformations (Salmon et al. 2016). One such client protein is Im7, an 87 amino acids single domain protein, that inhibits the bacterial toxicity of its cognate protein, E7. In the absence of colicin E7, several residues located at the Im7 surface, like the two tyrosine residues at positions 55 and 56, are highly frustrated with increased flexibility (Ferreiro, Komives, and Wolynes 2014). These frustrated residues are responsible for the emergence of an on-pathway folding intermediate state that is stabilized by non-native interactions (Sutto et al. 2007). This high frustration at the native state of Im7 is released when the Im7-E7 complex is formed because Im7 has evolved to have both a significantly funneled landscape and a funneled binding landscape. The evolutionary importance of these highly frustrated residues in Im7 is evidenced by their conservation both in sequence identity and high frustration across its homologs (Parra et al. 2024). Spy was shown to selectively recognise and interact with those highly frustrated residues leaving the rest of the protein largely intact (He et al. 2016). As a consequence of the interaction with Spy the complete secondary and tertiary structure of Im7 is perturbed towards a more dynamic conformational ensemble.

## 6.2 Chaperonin GroE

The GroE chaperonin system in *E. Coli*, constituted by the GroEL and GroES complexes, facilitates ATP-dependent folding both *in vivo* and *in vitro* (Hayer-Hartl, Bracher, and Hartl 2016). GroEL consists of 14 identical units organized as two heptameric stacked rings with a cavity at each end (Braig et al. 1994). GroES, an homoheptameric single ring, caps the GroEL cavity and forms a compartment where folding can occur in isolation from the bulk solution. Intense efforts have been dedicated to understanding how GroEL assists folding and what distinguishes GroE substrates from other *E. Coli* proteins (Azia, Unger, and Horovitz 2012). *In vivo* experiments have shown that fewer than 100 proteins require folding assistance by GroE. These are referred to as obligate substrates. Bandyopadhyay and colleagues performed mutagenesis experiments (1-3 random mutations, and a designed variant with 11 mutations) on the enhanced green fluorescent protein (eGFP) which would affect its GroEL-dependence for folding compared to its wild-type version (Bandyopadhyay et al. 2017). They found that the changes in GroEL dependence arising from mutations could be explained by changes in frustration patterns. Mutations that decrease GroEL dependence decrease the frustration at what originally were highly frustrated residues with the effect of decreasing frustration. Conversely, mutations leading to an increased GroEL dependence would occur at already minimally frustrated residues. The increased frustration in those regions now makes GroEL assistance necessary. This suggests that GroEL dependence is related to the occurrence of highly frustrated patches and that GroEL itself recognises such patterns. Apparently, residues that are important for function on GroE substrates conflict with folding stability and such effect is compensated by binding to the GroEL.

A similar mechanism, studied with coarse-grained molecular dynamics simulations, has been suggested to be utilized to recruit and load client proteins into the Hsp90-Hsp70-Hop chaperone system (Verkhivker 2022).

## 7. DISULFIDE BONDS: STRUCTURE STABILIZERS OR FRUSTRATION INDUCERS?

Some small bioactive peptides lack a sufficient number of residues to form a hydrophobic core able to provide sufficient stability to fold. Disulfide bridges potentially can compensate for this lack of stability. An example of this is MCoTI-II (Felizmenio-Quimio, Daly, and Craik 2001), a 34-residue trypsin inhibitor from *Momordica cochinchinensis* a member of the cyclotide family, characterized by a cyclic cystine knot (CCK) structural motif. MCoTI-II's CCK motif involves three disulfide bridges that connect Cys4-Cys21, Cys11-Cys23, and Cys17-Cys29, coupled with a cyclic backbone. *In vivo* studies have shown that oxidation of the cysteines to form the cystine knot facilitates subsequent enzymatic cyclization. A study using all-atom molecular dynamics (MD) simulations (Yi Zhang et al. 2016) that started from both the folded and unfolded structures, with various different pairings of disulfide bridges shed light into how frustration comes into the folding picture. The simulations showed that the native state is somewhat frustrated and that the Cys11-Cys23 and Cys17-Cys29 are needed to fold the structure together. During folding simulations the Cys11-Cys23 and Cys17-Cys29 inter-residue distances are anticorrelated, i.e. when one bridge forms the other gets looser, indicating thus that the two bridges frustrate each other. This anticorrelation persists until the structure reaches its native state. The Cys4-Cys21 bridge is not needed for reaching the native structure and is seen to partially correlate with both of the other bridges. The protein lacking the Cys4-Cys21 bridge is somewhat more flexible, with Cys4 being in a highly conserved region among MCoTI-II's homologs. This suggests that the increased flexibility in the Cys4-Cys21 bridge might be of functional importance, but it is not clear exactly how. In this case, disulfide bridges seem to be able to bring together regions of the peptide that otherwise would not stay together due to highly frustrated interactions.

A different story has been described by Steltz and collabs (Stelzl et al. 2020) where they have shown that disulfide bonds are not used by proteins to stabilize frustrated regions but sometimes introduce energetic conflicts instead. They studied how the oxidation of two cysteine thiols to a disulfide bond, during the catalytic cycle of the N-terminal domain of the oxidoreductase DsbD (nDSBd), introduces frustration that translates into functional signals. The catalytic site of nDSBd is located at one end of its structure with the two active site cysteines (C103-C109) located on opposite sides of a beta-hairpin. The nDSBd can exist in an oxidized or reduced form, i.e. with or without a disulfide bond between the two cysteines. The cap-loop region that spans residues D68-E69-F70-Y71-G72, shields the cysteines by adopting a closed conformation in both states. When nDSBd is in complex with its binding partners, the cap-loop is found in an open conformation that allows interactions between the cysteine pairs that belong to the different interactors. By applying NMR experiments and Molecular Dynamics simulations it was shown that the disulfide bond between the Cys103-Cys109 cysteines in the oxidized form of nDSBd introduces frustration in the Phe70 residue that alternates between a gauche and trans conformations. It is this frustration in Phe70 that allows the cap-loop and the active site in nDsbD$_{ox}$ to sample open bound-like conformations in the absence of binding partners, favoring the interactions. Structural comparisons of the oxidized and reduced forms of nDSBd in *Neisseria meningitidis* suggest that this mechanism seems to be conserved across many Gram-negative bacterial species.

A similar mechanism has recently been described for the Tsa1 protein. Tsa1 is a yeast peroxiredoxin enzyme (PRDX) that plays a crucial role in protecting yeast cells from oxidative stress by reducing hydrogen peroxide and organic hydroperoxides. Tsai1 is a typical 2-Cys peroxiredoxin that contains two conserved cysteines that participate in the catalytic cycle. The peroxidatic cysteine (Cp, Cys47) is responsible for the catalytic activity which reacts with a peroxide (ROOH) reducing it to its non toxic form (ROH) and oxidizing itself in the process to a sulfenic acid (cp-SOH). Afterwards, the Cp is regenerated by forming a disulfide bond with a so-called regenerative cysteine (Cr, Cys170). The disulfide bond Cp-S-S-Cr is then reduced by the NADPH-dependent Trx/TrxR system (Hall et al. 2009). During these processes, the protein undergoes a variety of structural modifications. In the oxidized form, large regions cannot be modeled in the crystal structure indicating the presence of significant structural disorder. Tsa1 switches between different oligomeric states (dimers, decamers or higher order ones) according to conditions such as the redox state or pH. In addition, PRDXs are able to work as chaperones, preventing other proteins from aggregation. The peroxidase function is predominant in the lower molecular weight forms, whereas the chaperone function is more common in the higher molecular weight complexes. It has been shown that oxidative stress and heat shock exposure of yeasts causes the shift from low molecular weight species to high molecular weight complexes which thus triggers a peroxidase-to-chaperone functional switch. These changes are primarily guided by the Cys(47), which functions as an efficient "H(2)O(2)-sensor" in the cells. The direct link between the structural and dynamic consequences of disulfide formation and oligomerization modulates the shift between the peroxidase and chaperone functions. A recent study (Troussicot et al., 2023) suggests that the oxidized form, that contains the disulfide bond between the two cysteines, constraints locally the dynamics of the protein, preventing nearby favorable interactions to be formed which in turn leads to an increased conformational dynamics in adjacent secondary elements. This increased dynamics would lead to the exposure of hydrophobic patches, increasing the conformation entropy (Troussicot et al., 2023). The exposure of hydrophobic patches and the increased conformational entropy favored by the disulfide formation would be linked with the chaperone function.

A different scenario is described in a recent study where Capdevila and collaborators have analyzed the structural basis for the persulfide-sensing sensitivity in a transcriptional regulator, the sulfide-responsive transcriptional repressor SqrR (Capdevila et al. 2021). In their work, they studied the chemical specificity of SqrR towards organic and inorganic sulfane (sulfur-bonded) sulfur species. They evaluated five different crystallographic structures of SqrR in various derivatized states and found that persulfide selectivity may be determined by structural frustration of the disulfide form that favors formation of the tetrasulfide product. In this case the functional tetrasulfide structure between Cys41 and Cys107 is favored due to its lower degree of highly frustrated interactions over the disulfide bridge that could be formed between them. The disulfide structure is selected negatively because it introduces structural frustration that results in increased conformational entropy in nearby regions that do not allow the protein to work properly. This case is inverse to the one reported by Steltz and Zheng.

Finally, Azurin from Pseudomonas aeruginosa is a 128-residue long protein that is constituted by eight β-strands, two α-helices, and two 310- helices arranged in a β-barrel topology that contains a blue copper site that facilitates electron transfer. When folded, Azurin contains a disulfide bond in its N-terminal region involving Cys3 and Cys26 that forms a closed loop (also named zero order loop), also known as the cinch including **β**1 and **β**2. There are no large changes on the protein folded state, in its apo form, when the copper ion is removed. Mutational studies *in vitro* have shown that replacing Cys3 and Cys26 with Ser or Ala side chains which eliminate the disulfide bond, decrease the stability of folded azurin (Dombkowski, Sultana, and Craig 2014). Although it has been shown that thermal and chemical stability are dramatically reduced *in vitro* when the disulfide bond is absent (Guzzi et al. 1999) the protein adopts a folded structure identical to that with an intact disulfide (Bonander et al. 2000). The covalent loop that results from the Cys3-Cys26 disulfide bond in azurin shortens the protein length by approximately 20% and restricts contact formation of residues within the covalent loop with other secondary structure elements. Disulfide bonds in proteins create closed loops in polypeptide chains, restricting protein conformational dynamics which would help to maintain protein structure and their stability (Barford 2004). A recent study by Zegarra and coworkers (Zegarra et al. 2021) focused on analyzing the effect of the zero order loop in the folding/unfolding of Apozurin (without the copper ion). The authors analyzed the folding/unfolding of the protein in models with and without the disulfide bond using coarse grained MD simulations. Their study found that the presence of the zero order loop not only increases protein stability but it also modulates the transition state ensemble. They found that either removing the disulfide bond or by including the copper ion resulted in increased local frustration, distant from both the disulfide and the metal site locations. These frustration changes in turn affect the folding pathway although the folded state remains the same. Removing the disulfide bond results in the appearance of a new transition state ensemble core in another part of the protein. This study suggests that the zero order loop could help to stabilize certain folding intermediates over others facilitating a faster folding process.

Azurin is a prime example where local frustration has been linked to chemical activity. Long-range electron tunneling through metalloproteins is facilitated by evolutionary tuning of donor–acceptor electronic couplings, formal electrochemical potentials, and active-site reorganization energies. Although the minimal frustration of the folding landscape enables this tuning, residual frustration in the vicinity of the metallocofactor can allow conformational fluctuations required for protein function. Chen *et al.* showed that the constrained copper site in wild-type azurin is governed by an intricate pattern of minimally frustrated local and distant interactions that together enable rapid electron flow to and from the protein (Chen et al. 2022). In contrast, sluggish electron transfer reactions (unfavorable reorganization energies) of active-site azurin variants are attributable to increased frustration near to as well as distant from the copper site, along with an exaggerated oxidation-state dependence of both minimally and highly frustrated interaction patterns (Zong et al. 2007).

## 8. CONFORMATIONAL DIVERSITY AND LOCAL FRUSTRATION

As mentioned at the introduction, proteins fold by minimizing the internal conflicts between the aminoacids that compose their polypeptidic chain, satisfying the minimum frustration principle. But even at their folded state, not all conflicts can be minimised and some interactions remain in conflict, they are frustrated. The native state ensemble of a protein is composed of many different conformations, of similar energy, that exist in a dynamic equilibrium. Local frustration drives the interconversion between such conformations at the bottom of the folding energy landscape. Different studies have explored the relationship between local energetic frustration, the folding process of proteins and the conformational diversity of their native states.

**8.1 Local frustration analysis across molecular dynamics trajectories**

In 2021, Rausch et al (Rausch et al. 2021), developed a new version of the frustratometeR tool that included new functions to evaluate the impact of point mutations as well as to analyse the changes of frustration as a function of time from molecular dynamics simulations. To show how this new tool can help to understand the role of frustration through the folding process of a protein, they analysed the case of the IκBα protein, a inhibitor of NFκB. Analysis IκBα revealed significant differences in local frustration when the protein was studied as an isolated monomer (Fig. 4A) versus as part of its quaternary complex with NFκB (Fig.4B). In the monomeric state, certain regions of IκBα displayed higher frustration due to the absence of compensatory interactions provided by its binding partner (Fig. 4C vs. Fig. 4D). These findings underscore the dynamic nature of frustration and its dependency on the protein's context within a biological system. Exploiting these differences can offer insights into the mechanisms of protein-protein interactions and their evolutionary optimization.

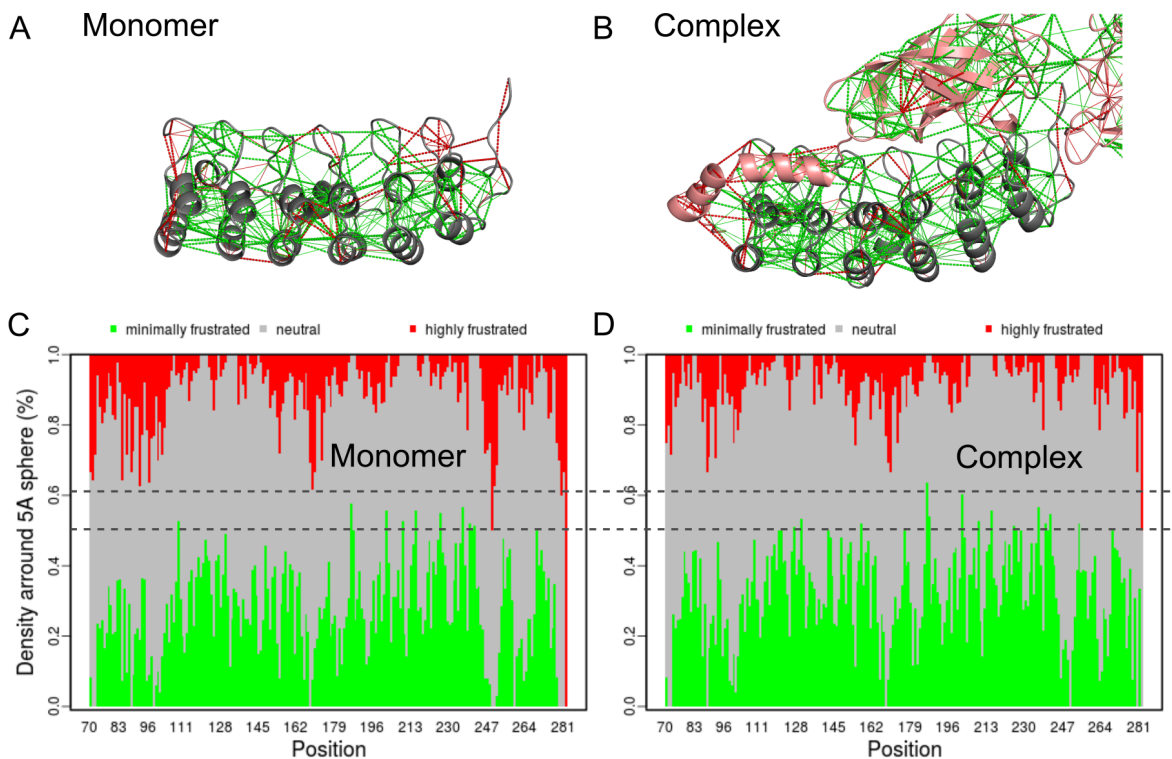

**Figure 4.   Differences in frustration in IκBα monomer vs complex.** Frustration was calculated for IκBα using the structure in the PDB=1NFI. **A.** Frustration results using as input only the chain that corresponds to IκBα (1NFI chain E). IκBα is represented in cartoon and coloured gray. Highly frustrated contacts are visualised in red and minimally frustrated contacts are visualised in green. **B.** Frustration results using as input the chains that correspond to IκBα in complex with NFκB (1NFI chains E and C). NfκB is coloured in salmon. Highly frustrated contacts are visualised in red and minimally frustrated contacts are visualised in green. **C.** Proportions of frustration contact types around each residue of IκBα as a monomer. **D.** Proportions of frustration contact types around each residue of IκBα as a complex.

To deepen the analysis the authors simulated the folding process of IκBα using a structure based coarse-grained model where the protein is unfolded and forced to gradually fold back into the native structure. Projections of the folding energy of the different conformations on a two dimensional space composed by two global coordinates, the fraction of native contacts Qw and the radius of gyration Rg. This two dimensional energy landscape representation facilitates the identification of the folded (F) and unfolded (U) states as well as a folding intermediate state (I) and two expanded states (E1 and E2) (Fig. 5A). Because this is a biased simulation, Qw is directly proportional to the simulation time. Several frames were sampled from the folding trajectory for which frustration was calculated. Variable residues, in frustration terms, were identified based on their average frustration values and the dynamic range across simulation frames. Frustration values of the variable residues across the entire simulation were used as input to perform a Principal Component Analysis (PCA) to reduce the dimensionality of the data and capture key trends in frustration dynamics. Residue pairwise correlations across the most important PCA components were then calculated to construct a similarity network that would facilitate grouping together those residues with similar frustration dynamic profiles. The network was clustered to identify modules of residues with similar behaviors (Fig. 5B). It was interesting to observe that indeed residues belonging to the same cluster displayed similar frustration dynamics across the simulation. Strikingly, 6 of the 7 clusters display sharp changes of frustration during the simulation. These changes happen at two specific moments that coincide with the two folding transitions, that were identified by the folding simulation and have experimental support, associated with the mechanism by which IκBα recognises and folds upon NFκB. A first group of residues has its major frustration change at a Qw that coincides with the occurrence of the U to I transition (Fig. 5C). A second group of residues has the mentioned frustration change at a Qw that coincides with the I to E1 transition (Fig. 5D). Furthermore, within both groups there are residues that increase their local frustration over time (given that the frustration calculations were made in absence of the NFκB protein) that correspond to quaternary interactions between IκBα and NFκB while others decrease their frustration, corresponding to those residues involved in tertiary interaction within IκBα. This is a clear example of how analyzing local frustration in combination with molecular dynamics simulations can help to better understand the folding mechanisms of proteins.

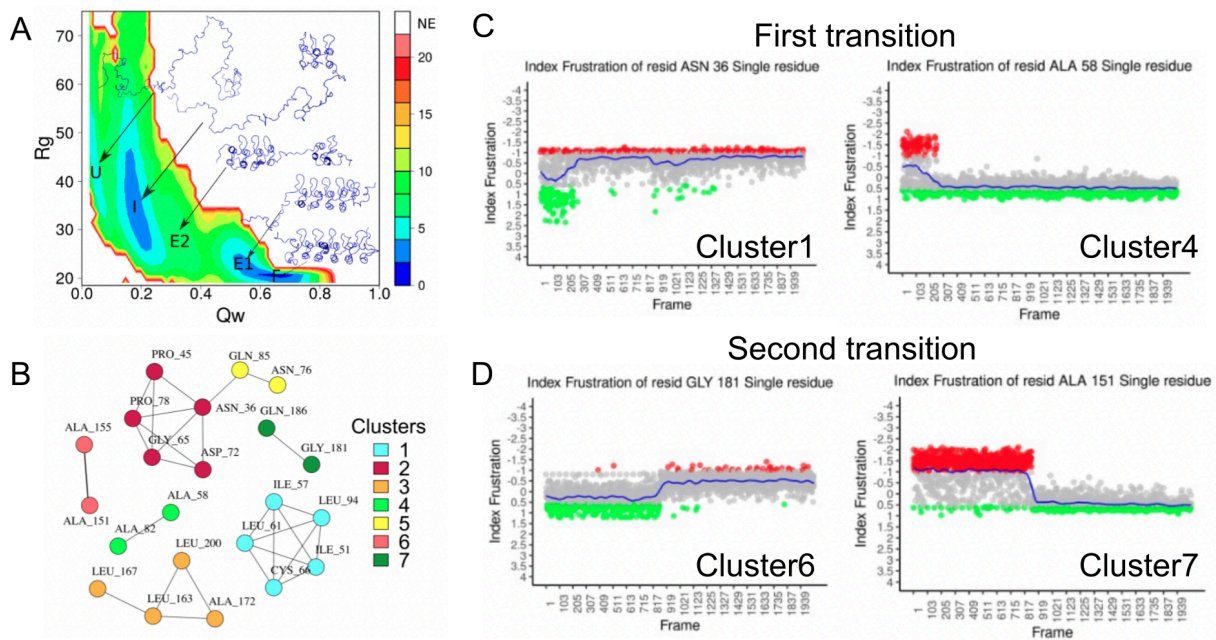

**Figure 5. Local frustration analysis during the folding process of IκBα. A.** Folding energy projections as a function of the folding coordinate Qw and the radius of gyration Rg. **B.** Correlation similarity network of frustration profiles across the simulation. Different colors correspond to different clusters in the network that group residues with similar frustration dynamics. **C.** Examples of two residues from cluster 1 and 4 where the major frustration change occurs at the location of the folding transition from U to I. Each point represents the frustration value for the residue for each frame, plotted as a function of the frame order (correlated with Qw). Points are coloured according to the frustration class they belong to, i.e. minimally (green), neutral (gray) or highly frustrated (red). **D.** Examples of two residues from cluster 6 and 7 where the major frustration change occurs at the location of the folding transition from I to E1.

## 8.2 Local frustration analysis lead to the identification of residues important for large conformational changes

Another subject of study that is very relevant at the moment consists in describing the conformational dynamics of the native state of proteins. A type of proteins that receive increasing attention nowadays are the so called metamorphic proteins that consist in protein sequences that can adopt dramatically different conformations according to their context, usually linked to different functions or regulatory processes. This is the case of the RfaH protein, a bacterial elongation factor that undergoes a unique metamorphic transition between two structural conformations: an α-hairpin (αCTD) and a β-barrel (βCTD) (Artsimovitch and Landick 2002). This structural flexibility is key to its function, as the αCTD auto inhibits the protein by interacting with the N-terminal domain (NTD), preventing interactions with RNA polymerase until specific triggers induce the conformational switch. Using frustration conservation analysis with the FrustraEvo tool (Freiberger et al. 2023; Parra et al. 2024), several contacts were identified between the αCTD and NTD that stabilize the autoinhibited conformation (Fig. 6A). These contacts were minimally frustrated and conserved across RfaH homologs, suggesting their evolutionary role in maintaining foldability and stability. Nine residues were observed to concentrate most of these minimally frustrated interactions with the hypothesis of these being responsible for holding the αCTD conformation. Each of these residues was mutated to all possible alternative amino acid identities to predict changes in frustration upon mutation. In Fig. 6B, frustration changes upon mutation are shown for residue F51. It can be seen that mutating the phenylalanine to a leucine would result in similar frustration values in the interactions between the residue and its neighbors while doing so to a lysine would considerably increase frustration. Two groups of mutations were selected for these residues: a set of mutations that would preserve similar minimal frustration patterns (stabilizing the αCTD) and a second set that would maximize the introduction of high frustration (presumably, favoring the βCTD transition). Structure predictions using AlphaFold2 supported the hypothesis that introducing highly frustrated mutations at these residues could drive the αCTD-to-βCTD transition, mimicking the natural metamorphic mechanism. Furthermore, when comparing the sequence of RfaH with the one of NusG, a non-metamorphic homolog that only folds into the βCTD conformation from which RfaH is believed to descend from via gene duplication divergence (B. Wang et al. 2020) , 6 of the analysed 9 residues change their identities and only one, L142S increases frustration in the αCTD conformation. Introducing the 6 NusG identities into RfaH makes AlphaFold2 to predict a βCTD confirmation in RfaH (Fig. 6C). Even more, introducing only the L142S mutation in RfaH, already forces AlphaFold2 to predict a structure of the CTD, that although not entirely similar to the native βCTD, contains high degrees of β secondary structure elements (Fig. 6D), suggesting the importance of the L142S residue to regulate the conformational change. In a follow up study, (González-Higueras et al. 2024) the same authors performed molecular dynamics simulations that capture the conformational changes that occur during the interconversion between the αCTD and βCTD and by applying a similar methodology as the one described for the IκBα/NFκB example confirmed the importance of L142 for the conformational transition detecting at the same time other residues that are important to it. Some of these residues have been previously reported by experimental studies while others are new propositions to be tested.

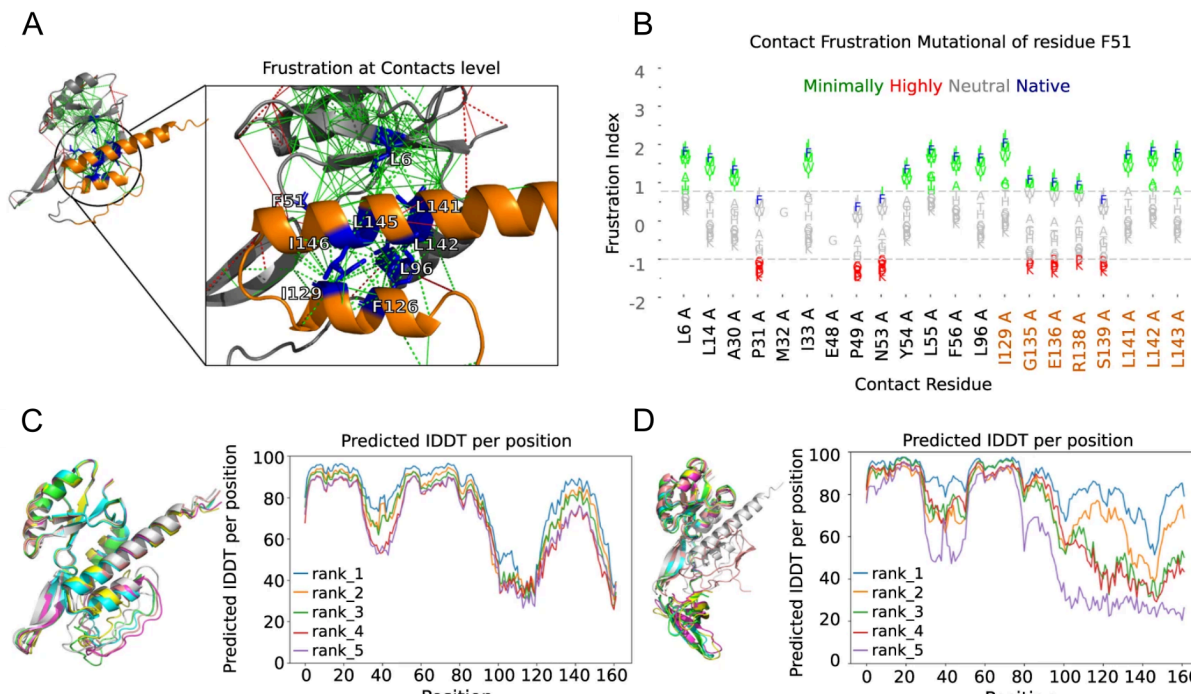

**Figure 6: A.** FrustraEvo Mutational index results. Red lines correspond to highly frustrated interactions, and green lines to minimally frustrated interactions. Orange backbone corresponds to the interdomain region (CTD), and residues in blue and sticks are the nine interface residues. **B.** Frustration changes upon mutation for Phe 51 using FrustratometeR. The x axis shows the residues with which the residue, either wild-type (Phe) or mutated, establishes contacts in the structure. In the y axis, we show the mutational frustration index for the contacts. The wild-type amino-acid identity is shown in blue, and the variants are colored according to their frustration state. **C.** AlphaFold2 top five predicted models superimposed for RfaH containing different sets of mutations with frustration values similar to the wild type and **D.** having mutations that maximise the increase of frustration compared to the wild type. Figure reproduced from (Freiberger et al. 2023), licensed under the Creative Commons Attribution 4.0 International License (http://creativecommons.org/licenses/by/4.0/).

## 8.3 Local frustration analysis allow AlphaFold2 to predict conformational motions:

The case of metamorphic proteins represents an extreme example of conformational variability of a protein native state. However, many globular proteins have structurally distinct states that although not entirely conformationally different do display key differences that are important for function. This behaviour is seen in the case of the active versus inactive conformations of kinases and the tense (T) versus relaxed (R) states in Heamoglobin. A recent study (Guan et al. 2024) presents an innovative approach for predicting protein conformational motions by integrating AlphaFold2 with energetic frustration analysis. Without using frustration analysis, AlphaFold2 is able to accurately predict a structure of a protein although it does this in a static way. In the case of proteins whose function is associated to the interconversion between different conformations, a priori it would not seem clear why AlphaFold2 would favour one over the other(s) although all of them need to be somehow encoded in the sequence. As an example of how local frustration can be useful to study conformational diversity in proteins Guan et al. used the Adenylate kinase (AdK) protein. AdK is a key protein to regulate the cellular adenosine triphosphate (ATP) levels by reversibly catalysing the phosphate transfer between ATP and adenosine monophosphate to produce two adenosine diphosphate molecules (Dzeja and Terzic 2009). During the catalytic cycle AdK exhibits large conformational transitions between open and closed conformations going through several intermediate structures (Henzler-Wildman et al. 2007). Analysis of the local frustration patterns in the open and closed conformations revealed that the domain interfaces of the closed structure of AdK are enriched in contacts with high frustration which might facilitate the conformational change towards the open structure. When predicting the structure of the AdK, AlphaFold2 returns a model that closely resembles the closed conformation with the same high frustration enrichment as the crystal structure. Other studies have shown that different conformations can be obtained if the input MSAs used by AlphaFold2 are manipulated, e.g. by clustering the sequences and using sub MSAs as input (Wayment-Steele et al. 2024). Instead of clustering the MSA by sequence similarity, Guan et al. developed a strategy that clusters sequences according to the total frustration that would result if such sequences would fold into a given target structure (Fig. 7A). They threaded the sequences of the MSA into the AdK native structure and calculated the total energies for all the native contacts (ET) as well as for the highly frustrated contacts (EHF) using the Rosetta energy function. By doing that, it was observed that there were large differences in the energies calculated across homologs in the MSA. While some sequences displayed high frustration at the binding domain as the *E. Coli* AdK others did the opposite. The sequences can be grouped according to their energetic patterns in the ET - EHF landscape (Fig. 7B). If a sub MSA composed of the sequences that minimize both quantities, ET and EHF are used as input to AlphaFold2 the closed conformation is predicted (Fig. 7C). Conversely, if the top 50 sequences from the MSA with high ET and high EHF are used instead, the open conformation is predicted (Fig. 7D). Finally, one can introduce a parameter, in this case referred to as W, that fine tunes the mixture of sequences from each group of proteins, i.e with high or low frustration, that will be included in the input MSAs to AlphaFol2. Using a series of increasing values of W produces a continuous range of conformations between the open and closed conformations which resembles what has been observed in different experimental X-ray crystallographic structures (Fig. 7E).

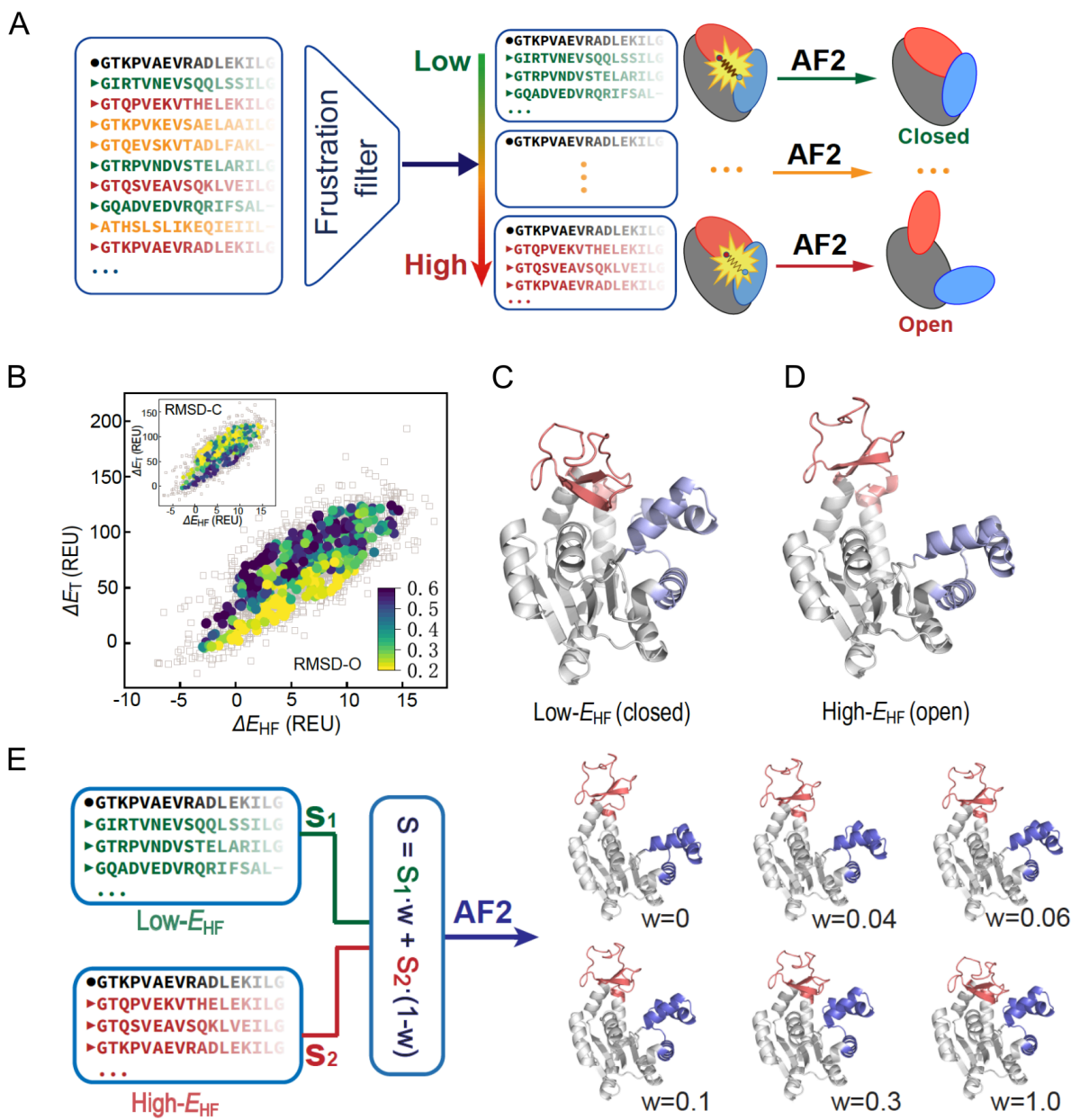

**Figure 7: Predicting conformational motions with AlphaFold2 and frustration analysis**. **A.** The workflow of the frustration-filtering strategy in predicting protein conformational motions. **B.** Two-dimensional space based on ET and EHF to select groups of sequences to be used as input to AlphaFold2 to force the prediction of different conformations. **C.** AlphaFold2 prediction when using low EHF sequences that lead to the prediction of a closed conformation. **D.** AlphaFold2 prediction when using high EHF sequences that lead to the prediction of an open conformation. **E.** Introducing the w parameter that generates mixtures with variable proportions of high and low EHF sequences lead to the prediction of a variety of conformational states between the open and closed conformation. Figure adapted from (Guan et al. 2024).

**9. LOCAL FRUSTRATION AND THE EVOLUTION OF PROTEINS:**

As mentioned before, proteins are very particular, evolved polymers whose sequences have been tuned so that both their structures and their dynamics enable their function. Achieving an appropriate tradeoff between structural stability and flexibility in proteins is a delicate proposition (Tokuriki and Tawfik 2009).

**9.1 Modularity in folding and evolution**

The concept of modularity in protein folding has a long history. Wetlaufer proposed that proteins might fold through parallel pathways involving independent folding units, a crucial idea for addressing Levinthal's paradox (Wetlaufer 1973). This notion gained further traction with the discovery of introns, leading Gilbert and Blake to hypothesize that exons, the protein-coding segments of genes, could correspond to independently folding protein units like domains or supersecondary structures (Gilbert 1978). This "exon-foldon" hypothesis suggests that the modular structure of genes reflects an intrinsic modularity of protein folding, with exons potentially representing evolutionary units that can shuffle and recombine to generate novel protein structures. Energy landscape theory provides a framework for understanding how proteins fold efficiently. Minimally frustrated proteins can fold through parallel pathways guided by native contacts. While not explicitly requiring pre-defined modules, the landscape may exhibit regions of preference, with some segments folding more independently due to strong internal interactions. These independently folding units have been termed "foldons" (Panchenko, Luthey-Schulten, and Wolynes 1996). Recent work by Galpern et al. supports the exon-foldon correspondence by demonstrating a correlation between conserved exons and segments with higher independent foldability as measured by a local frustration analysis (Galpern et al. 2024).

Repeat proteins, comprising ~20% of eukaryotic proteins, offer a compelling case study. These proteins consist of tandem arrays of similar amino acid sequences, resulting in elongated structures with repeating structural elements. While domain boundaries are less clear-cut in these proteins, their folding often involves cooperative interactions between neighboring repeats (Espada et al. 2015). The folding of repeat proteins is typically characterized by nucleation in one region and the propagation of structure to neighboring repeats (Sanches et al. 2022). Simple models demonstrate that parallel folding pathways can emerge, with the specific pathways that are most often taken, being influenced by the local energetics within each repeat. This "plasticity" of the energy landscape makes repeat proteins amenable to both experimental and computational studies. It was reported that the variations in the local energetics determine the appearance of a zoo of possible folding mechanisms, from two-state to multi-state to downhill scenarios (Galpern et al. 2022).

Intriguingly, exons in repeat proteins often encode single or multiple complete repeats, suggesting a strong link between gene structure and protein architecture. Studies on ankyrin repeat proteins have shown a correlation between sequence conservation and the conservation of local frustration patterns, further supporting the idea that conserved exons correspond to functionally and structurally important units within the protein (Parra et al. 2015).

## 9.2 Evolution and frustration adaptation of PLPro in Coronaviruses: An automatic strategy to study the evolution of viral variants

Recent advances in the implementation of machine learning for protein structure prediction make it possible to analyze a sizeable majority of the protein space that has evolved on our planet. To better understand how SARSCov2 has evolved, a recent study (Freiberger et al. 2023) carried out an automatic and unsupervised proteome-wide analysis on the SARS-CoV-2 virus in the context of the entire phylogeny of beta coronaviruses. For each of the 28 proteins that compose the proteome of the virus a homolog search was made within all known beta coronaviruses and defined functional families for each of them. Subsequently, structure predictions were made for all these proteins using Alphafold2 (Jumper et al. 2021).

The beta coronaviruses can be divided into four lineages: Sarbecoronaviridae, Nobecoviridae, Merbecoviridae and Embecoviridae. By analyzing the phylogenetic tree of beta coronaviruses a series of evolutionary steps that occurred through the evolution of this viral family were reconstructed. Cysteines involved in the coordination of a $Zn^{2+}$ ion, indispensable for the protein to function, are minimally frustrated in the entire family of beta coronaviruses, highlighting the importance of such energetic signatures. When focusing on the Sarbecoronaviridae family, it was observed that proteins in that evolutionary branch have a highly frustrated tryptophan at position 106. It is known that the tryptophan at that position enhances enzyme activity, in comparison to the version of the enzyme in Merbecoviruses, by stabilizing the catalytic triad (Patchett et al. 2021). In that same position Merbecoviruses have a leucine. It was observed that the enzymatic activity of PLPro in the MERS virus is enhanced if the Leu106Trp mutation is introduced (Lei et al. 2014). Finally, there are two extra substitutions in SARSCov2 that are different from the most common amino acids in the rest of Sarbecoviruses. Thr225 and Lys232 are highly frustrated in SARSCov2 while the analogous Val225 and Gln232 are neutral in Sarbecoviruses. These two positions are involved in the interaction between PLPro and  ISG15 and Ub host proteins that modulate the immune response to the virus. It is believed that these two substitutions improved the virus ability to evade the host immune system. Among Sabercoviruses, the highly frustrated

Thr225 is only present in SARS-CoV-2 and in RaTG13 (Zhou et al. 2020), the latter being a likely bat progenitor of the COVID-19 virus, suggesting that such mutation was acquired before the virus was able to jump to infect humans. However, when analyzing the rest of possible amino acids in Sabercoviruses at position 225, there are two bat-infecting viruses, Rc-o319 and bat-SL-CoVZXC21, that contain a methionine in position 225 that is even more highly frustrated than the Thr225 present in SARSCov2. There is no data about how that methionine affects the activity and efficiency of the enzyme and whether that may have an effect on the infectivity capabilities of the virus. However, as high frustration is often associated with functional features, if a member of the Sarbecoviridae family with a methionine in position 225 is found to infect humans, it should be carefully analysed as it could lead to a virus with enhanced infecting capabilities. Finally, Lys232 is unique to SARS-CoV-2 within the Sarbecovirus family, suggesting a recent gain of function event.

**9.3 Binding specificity fine tuning in Calmodulin (CaM):**

An early study that explicitly included the evolutionary dimension in frustration analysis used a combinatory analysis of protein sequence evolution and local energetic frustration to identify how calmodulin has balanced diversification during evolution (Swarnendu Tripathi et al. 2015). CaM is a multi-specific binding protein that plays a crucial role in intracellular $Ca^{2+}$ signaling by regulation of various target proteins. CaM has a remarkable structural plasticity that modulates its promiscuity for interacting with hundreds of different targets. It contains four EF-hand motifs for $Ca^{2+}$ binding that are separated in two lobes each of them containing a binding pocket that interacts with its target proteins (Meador, Means, and Quiocho 1993). The analysis of the evolutionary conservation of residues in CaM with the Evolutionary Tracer method (Mihalek, Res, and Lichtarge 2004) showed that it was related to the frustration patterns from 60 CaM/target complexes as calculated by the Frustratometer (Parra et al. 2016; Rausch et al. 2021). By comparing these distinct quantities it was possible to separate CaM residues into novel discrete classes to provide insights about how evolution optimised CaM sequence to balance its promiscuity and specificity features.

While Evolutionary Tracer can classify residues into conserved and not conserved, the Frustratometer classifies them into minimally frustrated, neutral or highly frustrated. These two dimensions can be used to subclassify the CaM residues into six discrete classes. From these, the most interesting classes are the following ones: 1) Minimally frustrated and conserved residues are localised at the helices composing the four EF-hands and are key to maintain stability. 2) Highly frustrated and conserved residues correspond to residues such as the Asp at the $Ca^{2+}$ binding loops which are important to coordinate the ion. It is important to note that, the (coarse grained) Frustratometer algorithm version that was used, excludes heteroatoms (like $Ca^{2+}$) and therefore the highly frustrated signal that is observed comes from the fact that the $Ca^{2+}$ ion is not in place to compensate the energetics of the Asp residues in those loci. One of the strengths of the coarse grained version of the tool is that it detects uncompensated regions in the structure of a protein and permits one to elaborate hypotheses about it, such as the binding of small ligands. If the "All atoms" frustratometer version (M. Chen et al. 2020) would be used instead, with the $Ca^{2+}$ ion in place, the highly frustrated interactions observed with the coarse grained version would now appear as minimally frustrated as they would be energetically compensated by the presence of the $Ca^{2+}$ ion. 3) Highly frustrated and not conserved residues could contribute to expanding protein function. This class of residues account for 3.3% of the total in CaM, e.g. Tyr99 is the most highly frustrated residue in humans although not evolutionarily conserved. The Tyr99 residue is substituted by a Leu in the baker's yeast protein and by a Phe in fruit fly or barley that lead to a decrease in frustration. Tyr99 is used as a substrate for phosphorylation by the protein tyrosine kinase. Modifications at this locus induces conformational changes that lead to altered CaM binding. 4) Minimally frustrated and non-conserved residues are typically located at the CaM-target binding interface and seem to be important for fine tuning the target specificity without altering the local protein scaffold.

Finally, the study showed that Met residues in CaM exhibit a wide range of frustration levels, making them critical for target-binding promiscuity. Met residues at different positions either remained neutral or became highly frustrated depending on the target protein. This variation allows CaM to bind to a variety of targets while preserving its overall structure. For example, Met124 shows significant changes in frustration depending on whether CaM is bound to a target or not. The Met124Leu replacement in lower eukaryotes leads to decreased frustration indicating that Met residues evolve to enhance binding flexibility at the expense of stability, facilitating CaM's multi-target interaction capabilities.

### 9.4 All Atoms Local Frustration And EGFR Specificity

We have mentioned in the previous section that the $Ca^{2+}$ ion binding site in the CaM protein appears as highly frustrated because the coarse-grain AWSEM frustratometer is blind to anything but the amino acids. If, in some way, the algorithm would be able to "see" the $Ca^{2+}$ ion, the interactions between the ion and the protein should be minimally frustrated. This is what was explored in the study that presented a new version of the frustratometer that has an all-atom resolution instead of a coarse-grained one as the original tool. The all-atom frustratometer quantifies local frustration using an atomistic force field, specifically the Rosetta energy function instead of the AWSEM-MD potential. Unlike the coarse-grained version, the all-atom frustratometer accounts for detailed molecular interactions, including ligand binding and protein-protein interfaces, making it particularly useful for studying drug specificity and biomolecular recognition.

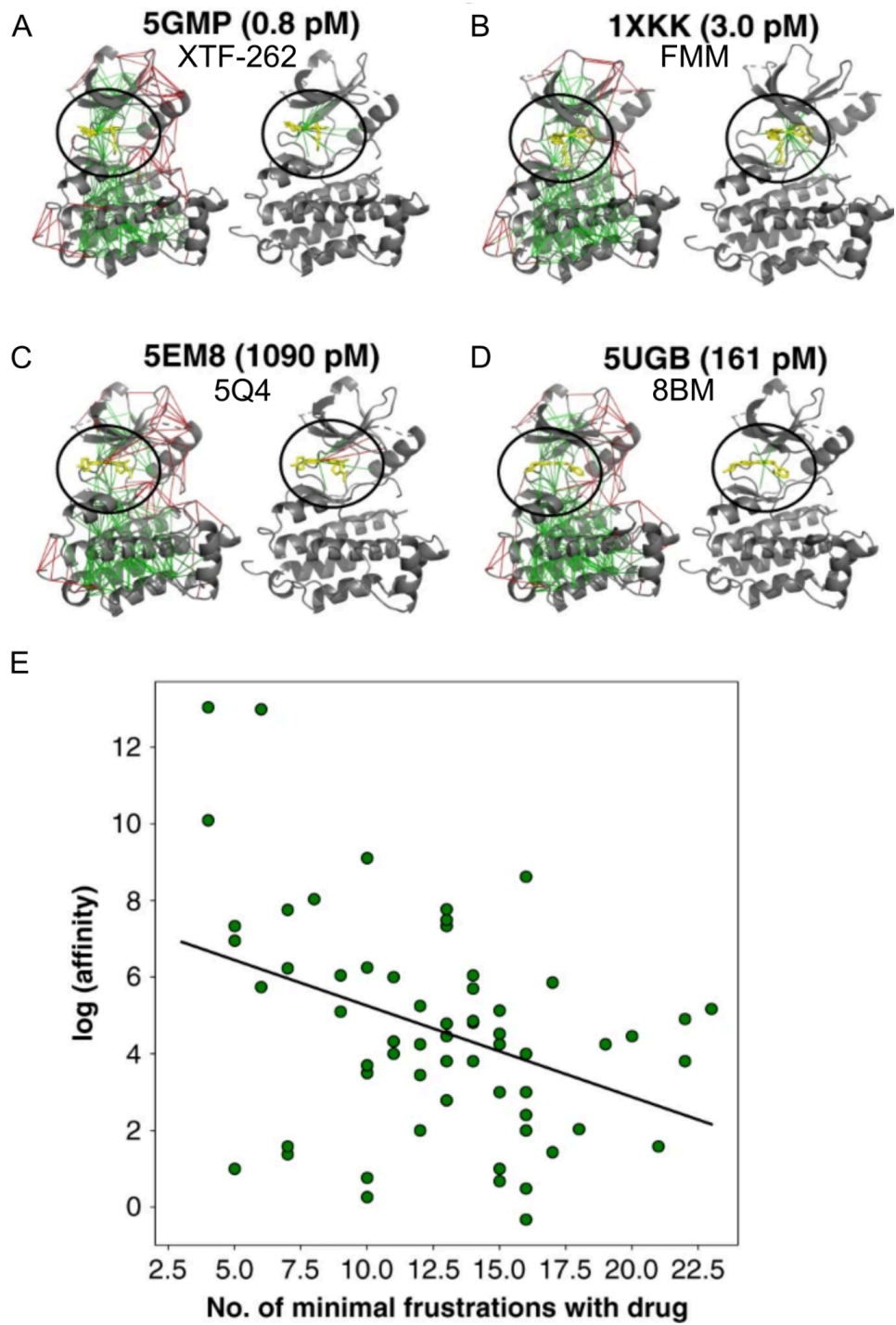

**Figure 8. Number of minimally frustrated interactions correlate with EGFR-inhibitors affinity.** Frustration patterns with four different inhibitors are show: **A.** XTF-262, B. FMM, **C.** 5Q4, **D.** 8BM. **E.** correlation between number of minimally frustrated interactions between the drugs and EFGR and the binding affinity in log scale. Figure is reproduced from (M.

The authors of the algorithm studied the relationship between localized frustration and drug binding specificity by analyzing different EGFR-inhibitor complexes. Using this new atomistic strategy, the authors examined the differences in the frustration patterns of different inhibitors when bound to the binding pocket of EGFR. The frustration patterns of EGFR in complexes with four different inhibitors are shown in Fig. 8A-D. XTF-262 forms more than ten minimally frustrating interactions with the pocket (Fig. 8A), while a weaker binder 5Q4 (Fig. 8C) forms only three minimally frustrated interactions and once bound even brings along with it the formation of additional highly frustrated interactions in the pocket. This suggests that strong inhibitors make multiple minimally frustrated interactions with the binding pocket, creating a stable and highly specific binding environment. In contrast, weaker inhibitors form fewer minimally frustrated interactions and even introduce additional highly frustrated contacts, suggesting a less optimal binding fit. This is supported by the moderate correlation (Fig. 8E) that is observed between the number of minimally frustrated interactions and the logarithm of binding affinity. All-atom frustration could be integrated into drug-design pipelines to discriminate between different candidates when bound to a specific target, as well as to improve existing ones by minimizing frustration at the target-drug interface.

## 10. LOCAL ENERGETIC FRUSTRATION CONSERVATION IN PROTEIN FAMILIES

Proteins with a common ancestor can be grouped into families and superfamilies, depending on the degree of divergence among them. Protein families are expanded by gene duplications and speciation events. At the gene level, members of the same protein family accumulate changes via processes such as point mutations, insertions or deletions. Multiple sequence alignments (MSAs) of protein families show that some positions are under strong evolutionary pressure, showing little variability, while others undergo neutral evolution (Bastolla, Roman, and Vendruscolo 1999). The neutrally evolving sites allow protein sequences to diffuse in sequence space while at the same time preserving the key stability, foldability and functional restrictions of the protein. By only looking at the sequence conservation patterns in MSAs it is not possible to assess which constraints are being preserved by evolutionary pressure. To overcome this, a strategy has been developed to analyse local frustration patterns in protein families by quantifying their conservation either at the level of single residues or pairwise interactions. In the following we describe several case studies where this strategy, that is now implemented as a tool named FrustraEvo (Parra et al. 2024), has been applied to better understand protein evolution, stability and function.

## 10.1 Local frustration conservation in the ANK family

The first family for which local frustration conservation was studied was the Ankyrin repeat protein family (ANKs) (Parra et al. 2015). This family consists of a diverse set of proteins, such as IκBα, Notch and P16, that contain a variable number of copies of a 33 residue long structural motif that is tandemly repeated, forming an alpha/beta solenoid structure. A distinctive feature of repeat proteins is that sequence can vary greatly among the repeats even beyond the limits of detectable homology (Espada et al. 2015). By exploiting the fact that structure diverges more slowly than sequences a protein tiling strategy was implemented (Parra et al. 2013) to separate 54 members of the ANK family into repeats for which an MSA could be built and thus local frustration conservation could be calculated for each of its columns.

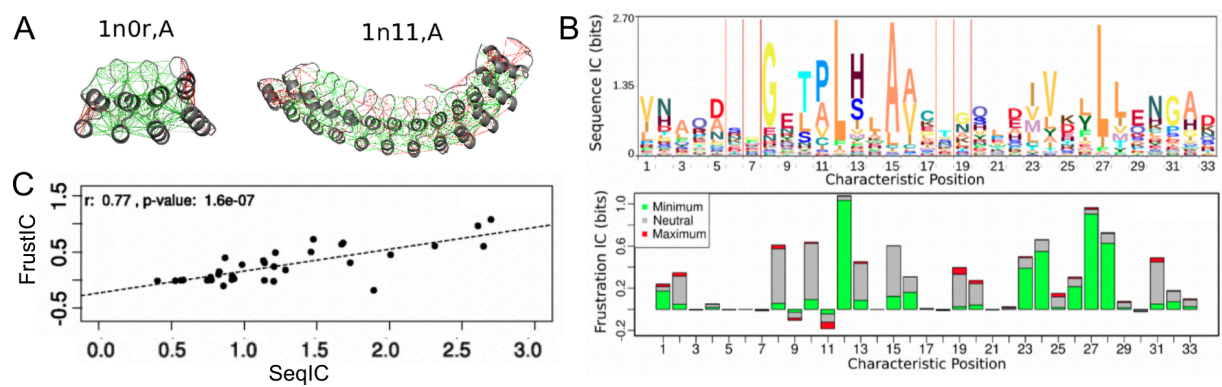

**Figure 9: Local frustration conservation in the ANK protein family. A.** Examples of two ANKs: 4ANK a designed ANK with 4 repeats and the D34 region containing 11 repeats from the AnkR a natural ANK. **B.** Sequence and local frustration logos showing sequence and frustration conservation across repeats extracted from a dataset of non redundant ANKs with experimental resolved structures. **C.** Correlation between sequence and frustration conservation measured as information content, SeqIC and FrustIC, respectively. Figure modified from (Parra et al. 2015) licensed under the Creative Commons international licence (Creative Commons Attribution License).

Local frustration was calculated for all members of the ANK family. Two examples of frustration for two ANK proteins with different numbers of repeats are shown in Fig. 9A. Sequences of all repeats were aligned to build an ANK repeat MSA and frustration values were mapped from the structures to the MSA. Conservation was quantified based on the Information Content (IC) derived from the distribution of both aminoacids and frustration states at each MSA column. This conservation of sequence and frustration can be represented in the form of sequence and frustration logos, respectively (Fig. 9B). Two observations emerge: 1) Although protein-protein interaction sites are enriched in highly frustrated interactions, only the minimally frustrated and the neutral positions were conserved across repeats in the MSA, 2) Frustration conservation (FrustIC) correlates with Sequence conservation (SeqIC) (Fig. 9C). The absence of conserved highly frustrated positions in the MSA can be understood as arising because the ANK family functions by interacting with a wide range of other proteins and such protein-protein interaction sites are not localized in alignable regions in the MSA. The correlation between FrustIC and SeqIC means that at each canonical position of the 33 residues that constitute a repeat, the consensus amino acid is the one that maximises folding stability. Interestingly, ANK proteins have proved amenable for consensus design (Mosavi, Minor, and Peng 2002) resulting in hyperstable proteins. These designs can later be evolved to bind specific partners by random mutations (Binz et al. 2003). Not surprisingly, the positions that are allowed to be varied when evolving designed ANKs correspond to positions with low FrustIC which can fine tune binding specificities without compromising structural integrity.

## 10.2 Local frustration conservation in catalytic families

In a follow up study, all enzymes from the Catalytic Site Atlas (CSA) with experimentally annotated catalytic residues were analyzed. It was observed that, collectively, catalytic sites and those involved in cofactors binding are enriched in highly frustrated interactions (Freiberger et al. 2019). Two enzymatic families, i.e. Beta-lactamases and Fructose-Bisphosphate Aldolases, were analyzed to study their local frustration conservation patterns. Sequence and frustration conservation were calculated for the family MSAs and represented as sequence and frustration logos. Results for the Beta-lactamases family are shown in Fig. 10A. In contrast to ANKs, where mostly neutral and minimally frustrated loci were conserved, in the enzymes conserved highly frustrated loci were also found which coincided with the location of several catalytic residues (Fig. 10B and 10C). These highly frustrated catalytic residues were also conserved at the sequence level. Moreover, when analysing contact mutational frustration, the catalytic residues were densely connected by highly frustrated interactions (Fig. 10D). The high conservation of sequence identity and high frustration of catalytic residues within these two families underscores their evolutionary importance and provides clear evidence of how function conflicts with stability in protein biophysics. Interestingly, in the Beta-Lactamases there were five non catalytic residues that are both conserved and highly frustrated. These loci are related to functional aspects such as substrate recognition and binding or the susceptibility to antibiotics. The latter suggested that the link between conserved high frustration and function in proteins is more general, going beyond catalysis alone. In these two enzymatic families, no correlation between FrustIC and SeqIC was observed as for ANKs.

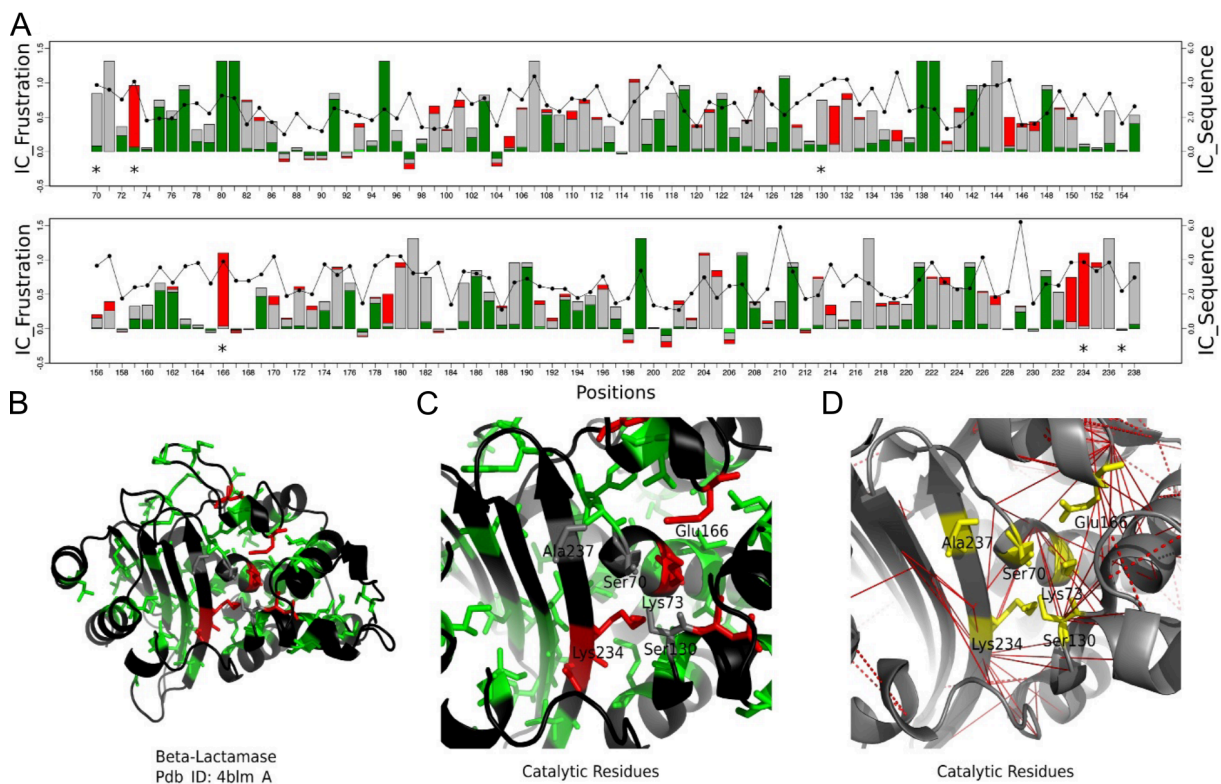

**Figure 10. Conservation of local frustration and sequence identity for the β-lactamases class A family. A.** Local frustration conservation (FrstIC) at the frustration single residue level. Green, minimally frustrated proportion; red, highly frustrated proportion; gray, neutral proportion. Black circles show the values of sequence conservation (SeqIC). **B.** β-Lactamase structure: in green and red, residues with FrstIC values greater than 0.5, minimally frustrated and highly frustrated, respectively. In black all other residues. **C.** Catalytic residues at the active site. **D.** Mutational frustration at the catalytic site. Red lines correspond to highly frustrated interactions of the catalytic residues (in yellow) among themselves and with nearby residues in the structure. Figure reproduced from (Freiberger et al. 2019)

## 10.3 Generalizing local frustration conservation analysis in protein families

From when the cases of Beta-lactamases and Aldolases were examined in 2019, it was initially challenging to expand the analysis due to the scarcity of structural data. It was extremely difficult at that moment to find protein families with more than 20 non-redundant members with experimentally solved structures. This changed in 2021 when AlphaFold2 (Jumper et al. 2021) was released and allowed scientists to generate good quality models of most known proteins, giving us the possibility to expand our analysis to thousands of protein families. Around that time, several reports were published about proteins for which double-deep protein fragment complementation (ddPCA) experiments have been performed (Faure et al. 2022; Weng et al. 2024). This experimental method was used to quantify the effects of amino acid variation on protein stability (abundance) and function (binding) (Faure et al. 2022; Weng et al. 2024). For each position of the C-terminal SH3 domain of the human growth factor receptor-bound protein 2 (GRB2-SH3), the third PDZ domain of the adaptor protein PSD95 (PSD95-PDZ3) and the GTPase KRas (KRAS), the experiments have produced two scores related to the effect of variations on stability and function. In order to test this strategy in a semi-supervised manner, homologs for each of these proteins were recovered by using blast and processed each dataset was processed by the S3Det algorithm (Rausell et al. 2010) that clusters the MSAs based on SDPs detection. Subsequently, the cluster containing each of the proteins of interest was taken, AlphaFold2 structural models were produced for each of the homologous proteins, and both the structures and the MSAs were used as inputs to calculate local frustration conservation patterns with FrustraEvo. GRB2-SH3 and PSD95-PDZ3, which are protein-protein interaction domains, show conservation only for minimally frustrated and neutral positions and a correlation between SeqIC and FrustIC resembling in this way what was observed for ANKs. In contrast, KRAS which is a globular protein as were Beta-Lactamases and Aldolases, shows conservation for neutral and minimally frustrated positions in addition to a highly frustrated and conserved lysine that is known to be important for recognising the nucleotide substrate. As in the case of the other two globular families, KRAS shows no significant correlation between SeqIC and FrustIC. After interpreting the examples described in these sections, it becomes clear that quantifying local frustration conservation in protein families can classify aligned residues in MSAs into 4 classes: 1) Conserved and minimally frustrated loci which are related to foldability and structural stability, 2) Conserved and neutral loci which serve as a buffer for conformational flexibility and evolvability, 3) Conserved and highly frustrated loci which are related to functional constraints and 4) Non energetically conserved loci that correspond to positions that are free to mutate among members of the same family without important consequences for either stability or for function. Finally the FrustIC values across positions in the MSAs of the families turned out to be relatively good predictors of changes in protein stability as measured by the ddPCA experimental method. It is remarkable that such a correlation exists even when frustration is a computational local descriptor of energetics while ddPCA is an experimental and global one.

## 10.4 Diversification of local frustration patterns in protein superfamilies

In the original paper where the algorithm for localizing frustration was presented (Ferreiro et al. 2007), it was reported that protein-protein interaction sites are enriched in highly frustrated interactions. This high frustration is released when the quaternary complexes are formed, meaning that the change in frustration between the monomers and the quaternary structures thermodynamically drives the protein-protein recognition process. Another molecule for which enough experimentally solved structures are available is Haemoglobin, a tetrameric complex composed of two copies of α-globin and two copies of β-globin. On top of its physiological importance related to oxygen transport, Haemoglobin is an interesting case study because the α and β globin families are both evolutionarily related and part of the same biological unit at the same time. When analysed a combined MSA of the two families, only neutral and minimally frustrated positions were conserved, pointing out to positions that are important for the overall fold integrity. In contrast, when each of the two families were analyzed separately by analysing family specific MSAs instead of a combined one, 12 highly frustrated and conserved positions were identified in α-globins and 8 in β-globins with only two being common to both lineages (Fig. 11A). Most of these positions correspond to residues that have been described as being relevant for protein-protein interactions involved in the tetrameric structure of Haemoglobin. Moreover, in the case of α-globin some of the highly frustrated positions correspond to residues that are recognized by AHSP, a chaperone that prevents α-globin from being toxic because of its aggregation propensity when it is not part of the tetrameric complex. Because α and β globins have a common ancestor within the globin superfamily, all observable amino acid differences between the two families are a consequence of divergent evolution. Some positions in the combined MSA between the two families show a mixture of two main amino acid identities, e.g. position 40 in the MSA that corresponds to Lys40α, that is conserved and highly frustrated in the α family, and Gln39β that is neutrally conserved in the β family. These positions, also known as Specificity Determining Positions (SDPs) account for positions that have been subjected to differential evolutionary pressure along the divergent evolutionary trajectories of the two families where the α family holds a functional requirement at those loci while there is only a neutral requirement in the β family. Thanks to the combination of ancestral sequence reconstruction and resurrection techniques (Thornton 2004) with the latest structure prediction algorithms such as AlphaFold2 (Jumper et al. 2021) the order of events that gave rise to the diversity observed within a superfamily like the globins can be reconstructed. This allowed the prediction of the structures of a set of ancestral sequences (Ancα, the ancestor of all α-globins; Ancβ the ancestor to all β globins; and AncMH the common ancestor to Myoglobins, α and β globins) reported in a study that explored the origins of complexity in Haemoglobin (Pillai et al. 2020). After structure prediction the frustration of each ancestral protein can be calculated and compared to the frustration patterns of extant globins. An analysis comparing the frustration values of a specific locus both at ancestral proteins and the ones of extant α and β globins, neuroglobins and myoglobins gives insights into the changes in evolution along the evolutionary history of the globin superfamily. When the local frustration of position 40 in the shared MSA (39α/40β according to each subfamily notation) was analyzed as a function of the phylogenetic depth of each protein (as a pseudotime coordinate), it was discovered that the ancestral versions of the globins contained a highly frustrated lysine, which was later replaced during evolution by other residues, leading to changes in frustration. β-globins substituted the ancestral lysine by a glutamic acid that becomes neutral while in Myoglobins and Neuroglobins the ancestral lysine is replaced by a minimally frustrated leucine. This change in Myoglobins and Neuroglobins can be explained as the residue in that position is not needed for protein-protein interactions as these two families are monomeric. This small computational experiment shows how this strategy can be used to better understand protein evolution to decipher how nature finely tunes protein sequences given the functional context of different subfamilies over evolutionary timescales. This approach could help us decipher how to rationally modify the properties of existing proteins, enabling their repurposing for other applications.

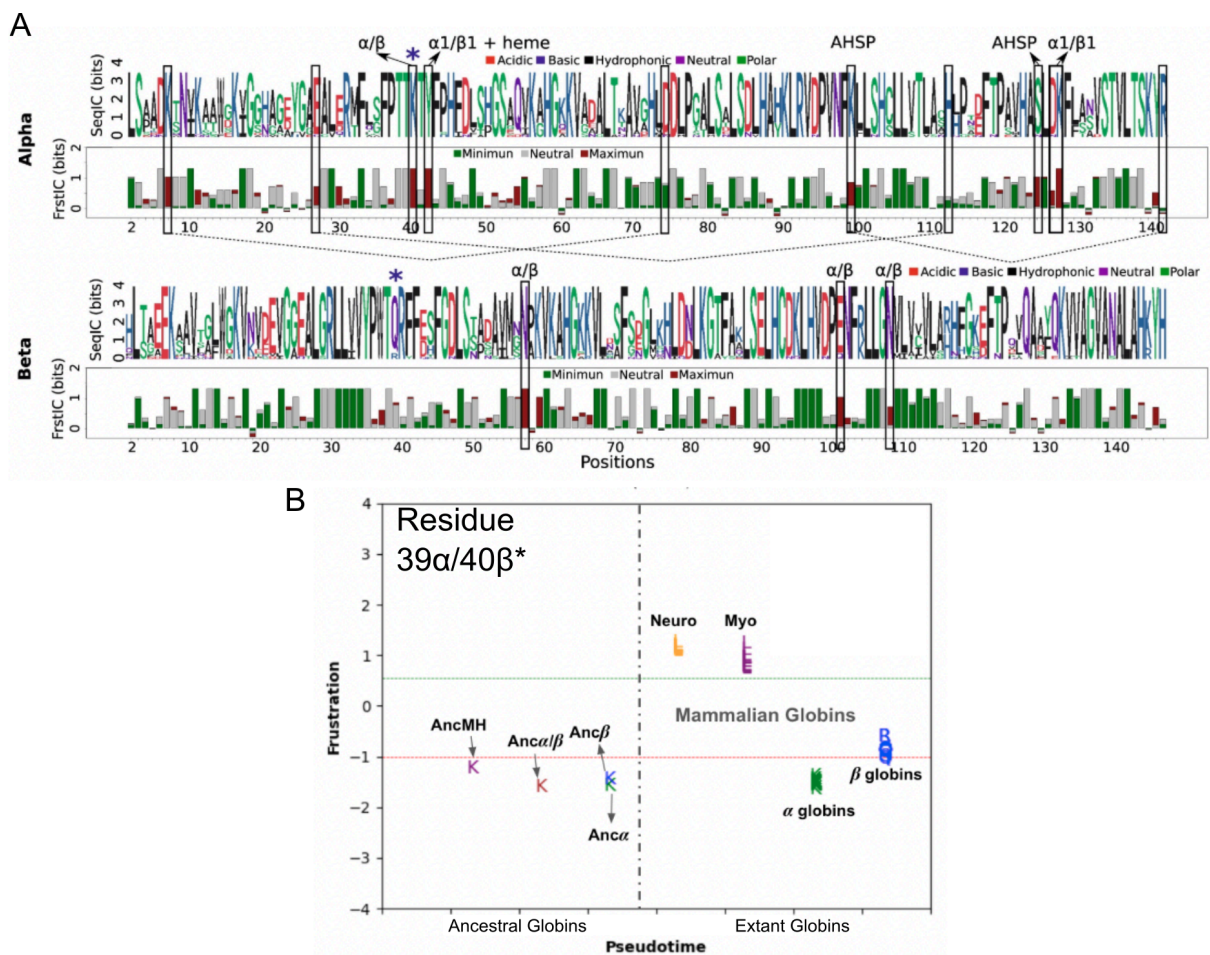

**Figure 11: Local frustration conservation divergence in the globin superfamily. A.** Local frustration conservation in α and β globins. Functional residues that are conserved and highly frustrated are marked with black rectangles and the associated protein-protein interactions are labeled. Figure panel adapted from (Freiberger et al. 2023) licensed under the Creative Commons Attribution 4.0 International License

(http://creativecommons.org/licenses/by/4.0/). B. Local frustration of residue 39α/40β in ancestral and extant globins. For ancestral proteins the x-axis orders proteins according to their ancestral age based on their phylogenetic depth; extant proteins are separated in the x-axis for visualization clarity. The y-axis shows the local frustration values for the residue marked with an asterisk in each family in panel A (position 40 in α globins, 39 in β globins) for each protein.

## 11. EXAMPLES OF FRUSTRATION IN PATHOLOGY

Proteins are dynamic molecules that live in an equilibrium between global stability and local instabilities to perform function by interacting with other proteins or different types of molecules. This delicate equilibrium can be dysregulated by the occurrence of mutations in the sequences of proteins that result in perturbations in their structures or modified dynamics. In certain cases these changes can lead to pathogenic phenotypes and the manifestation of diseases.

### 11.1 Frataxin and Friedreich ataxia

Frustration in proteins arises from conflicting requirements between folding and function. It occurs when it is impossible to simultaneously optimize all interactions among a system's components. While most globular proteins generally exhibit minimal frustration that ensures their rapid and reliable folding, functional constraints can introduce local frustration patterns, particularly in active sites. The neurodegenerative disease Friedreich ataxia results from a deficiency of frataxin, a mitochondrial protein. In frataxin, frustration analysis reveals there are non frustrated interactions in the β-sheet but that there are highly frustrated interactions within the α-helices and in their contacts between the α-helices and the β-sheet (Gianni et al. 2014). This frustration pattern correlates with the protein's function, involving multiple binding sites for metals and ferrochelatase. The presence of frustrated regions leads to a malleable transition state structure during folding, resulting in broad energy barriers and marked curvatures in the chevron plots which monitor how folding kinetics changes with stability through the addition of denaturants (Oliveberg and Wolynes 2005). Importantly, frataxin's sequence appears optimized to prevent aggregation in these frustrated, misfolding-prone regions, highlighting the complex interplay between frustration, function, folding, and aggregation prevention in protein evolution.

### 11.2 Cell cycle and cyclins

The progression of the cell cycle is regulated by cyclin proteins, which are often abnormally controlled in cancer cells. Researchers have examined the structural changes that occur during the activation of CDK/cyclin complexes by integrating various levels of resolution to efficiently explore the conformational space (Floquet et al. 2015). The transition between open and closed conformations was found to be described by the lowest-frequency normal modes, with this transition being facilitated by specific distributions of frustrated contacts within the complex. A recent review describes how the anisotropic nature of protein dynamics causes proteins to respond to external disturbances along only a limited number of intrinsic, large-amplitude directions (Kitao and Takemura 2017). Functional regulation of proteins can be achieved through changes in energetic frustration that occur as the molecule reconfigures along these large-amplitude motions owing to the frustration induced near degeneracy. These changes can be triggered by minor external perturbations, such as the binding of other molecules, resulting in the emergence of allosteric control.

### 11.3 Transthyretin

Frustration in protein folding refers to the nonoptimal free energy landscape that arises from conflicting functional constraints during the evolutionary process. In the context of the study of human transthyretin (TTR) and its ancestral enzyme, the *Escherichia coli* transthyretin-related protein (EcTRP), frustration plays a critical role in determining the stability and unfolding mechanisms of these proteins (Jäger, Kelly, and Gruebele 2024). EcTRP exhibits significant energetic frustration at its dimer–dimer interface, which is essential for its enzymatic activity. In contrast, TTR has evolved to minimize this frustration, thereby enhancing its kinetic stability as a hormone carrier. This reduction in frustration allows TTR to maintain its structural integrity and function effectively in the bloodstream, highlighting how evolutionary requirements determine the balance between protein stability and functionality. The presence of frustration in EcTRP, necessary for enzymatic activity, contributes to its sensitivity to mutations, whereas the low frustration of TTR correlates with its robust kinetic stability. This underscores how the intricate interplay between function and stability needs to be understood.

The presence of frustration in EcTRP is further confirmed by the mutational studies which demonstrate that specific residues, such as tyrosine 111, serve as gatekeepers of frustration. When the tyrosine residue is mutated to threonine, the resulting EcTRP variant exhibits reduced kinetic stability and altered unfolding behavior, indicating that frustration is intricately linked to the protein's functional constraints. In TTR, the absence of significant frustration allows for a more efficient binding of hormones, as the dimer–dimer interface is optimized for this function. The evolutionary trajectory from EcTRP to TTR illustrates how the reduction of frustration can facilitate the development of new binding functions while maintaining kinetic stability, underscoring the importance of frustration in the evolutionary adaptation of proteins (Jäger, Kelly, and Gruebele 2024).

**11.4 Aromatic amino acid decarboxylase (AADC) deficiency**

An example of this happens with the occurrence of certain mutations, like P330L, in the gene encoding the *aromatic amino acid decarboxylase (*AADC) enzyme, which plays a crucial role in synthesizing dopamine and serotonin (Bisello et al. 2022). This rare neurometabolic disease is monogenic and causes severe symptoms such as intellectual disability, movement disorders, and autonomic dysfunction. The P330L mutation in the catalytic loop of AADC does not result in significant changes in the enzyme's overall structure or the inability to bind its cofactor pyridoxal 5'-phosphate (PLP) but severely disrupts catalytic function. The catalytic loop of the protein can be positioned in either an open or a closed conformation (Fig. 12A). Specifically, P330 helps position the catalytic loop as a lid to close over the enzyme's active site, where catalysis takes place. The mutation significantly reduces the enzyme's catalytic efficiency by mispositioning of the catalytically competent external aldimine intermediate which results in reduced production of neurotransmitters. The authors of the study observed that the amount of highly frustrated interactions increases at the closed state of the catalytic loop when the P330L mutation occurs. When the mutation occurs, the configurational frustration index and the number of highly frustrated interactions stays the same in the closed configuration but the frustration in the open state decreases by one unfavorable interaction. Therefore, the P330L mutation leads to a less frustrated open catalytic loop. Proline normally imposes a conformational constraint, providing a "pivot" that helps the loop to close correctly over the active site. By introducing leucine, this mutation destabilizes the closed, catalytically active conformation, as indicated by a shift towards an open conformation. This shift correlates with the increased local frustration, impairing the catalytic efficiency of the enzyme and suggests that drugs targeting dopamine pathways might partially alleviate symptoms.

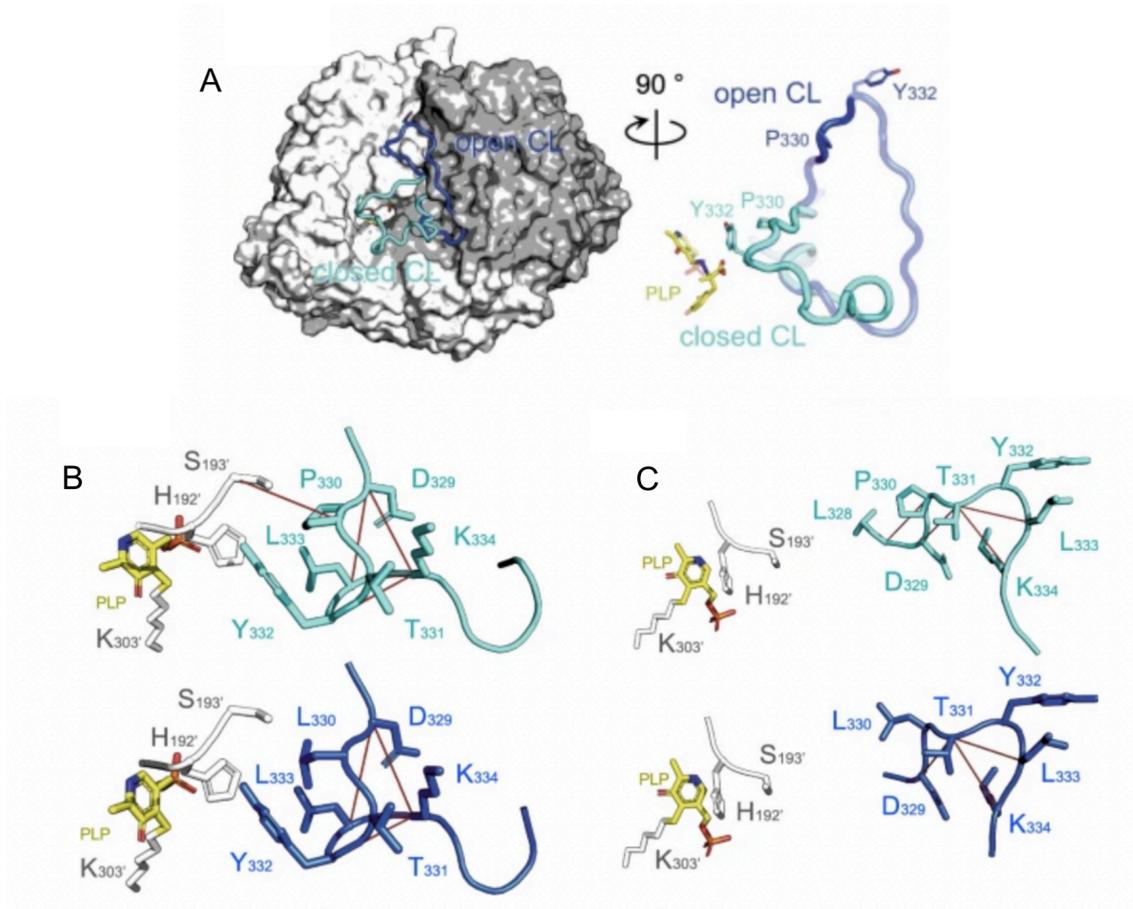

**Figure 12: Modeled catalytic loop (CL) of AADC. A** Dimeric AADC is shown as a surface with the two monomers colored white and gray. PLP, cDopa, Pro330 and Tyr332 are shown as sticks. The modeled closed catalytic loop (cartoon tube, cyan) of one monomer is inserted into the active site of the other monomer with Pro330 and Tyr332, both in close proximity to PLP of the facing monomer, while the modeled open catalytic loop (cartoon tube, blue) is completely solvent exposed. Frustration of the catalytic loop in both the closed **B** and open **C** conformation calculated for the WT (monomer A showing the catalytic loop in cyan, monomer B in white, PLP in yellow) and P330L (monomer A showing the catalytic loop in blue, monomer B in white, PLP in yellow) AADC. Conformational and mutational highly frustrated pairs are represented as red links on the same figure for clarity. The analyses were performed using the Frustratometer server that does not include coenzymes and for this reason PLP has been shown for clarity. As above, residues belonging to the opposite monomer of the catalytic loop are indicated with the prime. This figure has been adapted from (Bisello et al. 2022) and is licensed under a Creative Commons Attribution 4.0 international licence (http://creativecommons.org/licenses/by/4.0/). The original captions are used for the used panels.

## 11.5 Mutation in Desmin impairs proper polymerization

Desmin (DES) is a major intermediate filament protein crucial for the structural integrity and function of striated muscles. Mutations in DES have been associated with various forms of myopathies collectively known as "desminopathy." Recently, Castañeda et al, identified a novel heterozygous mutation (c.1059_1061dupGGA) in exon 6 of DES in an Argentine family with myofibrillar myopathy (Castañeda et al. 2023). This mutation leads to the duplication of a glutamic acid residue at position 353 (p.Glu353dup) of the DES protein. Clinical and myo-pathological evaluations of the index patient revealed characteristic features of myofibrillar myopathy, including muscle weakness, atrophy, and muscle fatty replacement. Local frustration analyses of DES dimer assembly revealed significant alterations in the coiled-coil structure and a more stable complex conformation when one or both monomers contain the mutation (Fig. 13A). The disease-causing mutation induces the formation of a minimally frustrated interface between the monomers, presumably stabilizing the dimer (Fig. 13B, 13C). The results suggest that p.Glu353dup mutation impairs the formation of a normal DES network after affecting its polymerization.

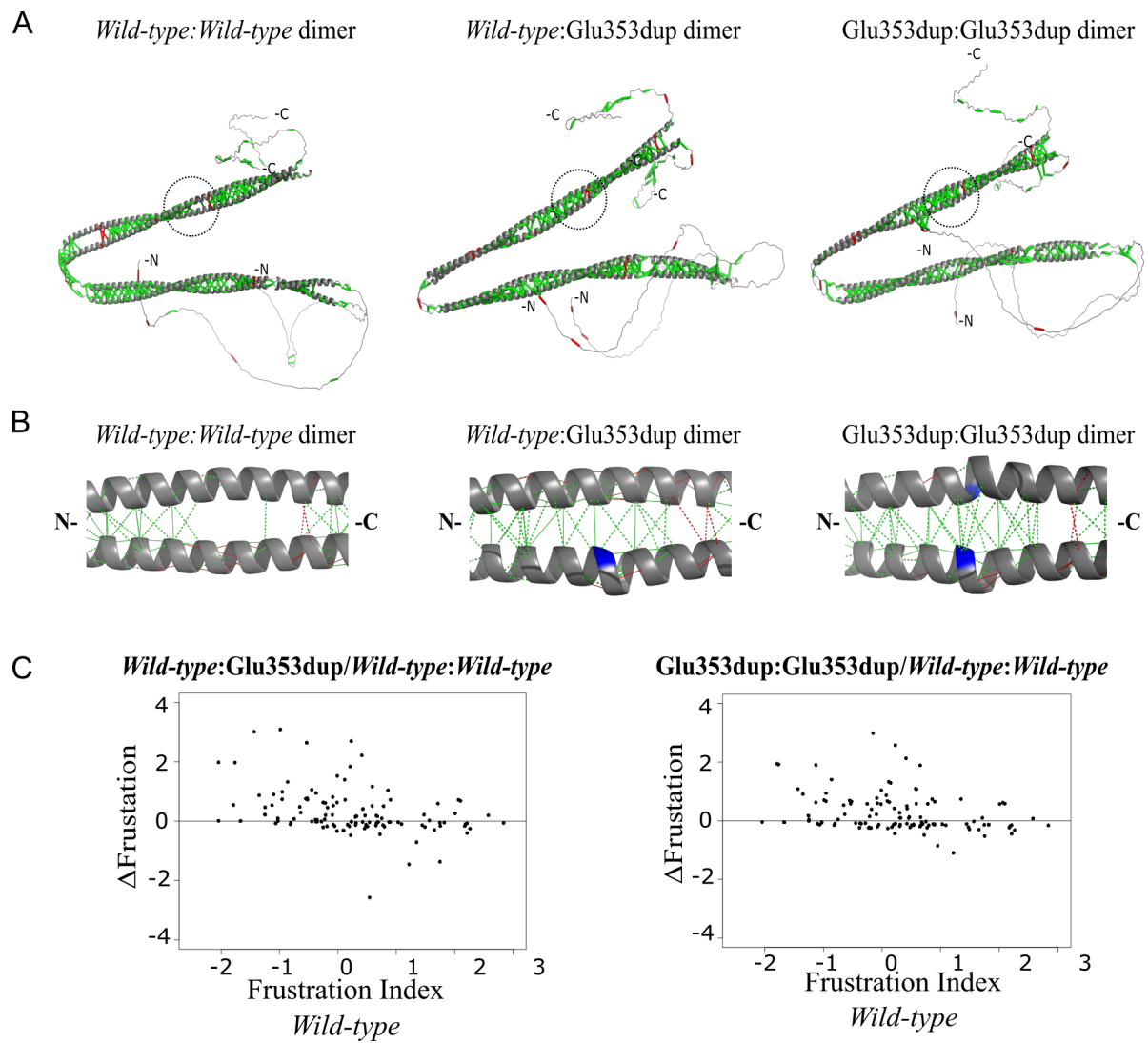

**Figure 13: Effect of Glu353dup mutation in DES dimer structure according to bioinformatic predictions.** Highly frustrated sites (Frustration index <= -1) are shown in red and minimally frustrated sites (Frustration index> 0.78) are shown in green. **A.** Comparison of alpha helix arrangement and amino acids interactions in DES dimer structures in presence and absence of the Glu353dup mutation. The mutation site is circled. It can be seen that the presence of Glu353dup mutation alters the periodicity of the supercoil, changing the alpha-helix crossing site and introduces more minimally frustrated contacts. **B.** Analysis of the frustration of the contacts in the area adjacent to the DES mutation site (shown in blue) in dimers that carry or do not carry the mutation. More minimally frustrated contacts are observed with the Glu353dup mutation. **C.** Evaluation of the change in frustration (delta Frustration) of each of the contacts between the dimers containing the mutation and the wild-type. Most of the contacts change the frustration level in a positive way, which means that they are more minimally frustrated. Figure reproduced from (Castañeda et al. 2023) licensed under a CC BY 4.0 License.

## 11.6 Mutations in p53

The Tumor Suppressor p53 is a protein that when malfunctioning leads to a drastic increase of diseases such as cancer. A recent study described how different mutations that affect p53 modified its conformational heterogeneity and local frustration features (Bizzarri 2022). The authors analyzed four different point mutations (p53R175H, p53R248Q, p53R273H, and p53R282W) which are known to affect the DNA binding and belong to the most frequent hot-spot mutations found in human cancers. p53 is a homo-tetrameric protein, where each monomer consists of an N-terminal transactivation domain (NTD), a C-terminal domain (CTD), and a large-structured region core domain (DBD) that binds to DNA response elements. It is worth noting that p53 is intrinsically disordered in its monomeric state. Differences in local frustration values of wildtype p53 (Fig. 14A) and the frustration values of the studied mutants (Fig. 14B) are compared in the following.

The mutation P53R175H, in which Arg175 is replaced by His, is one of the most common p53 mutants found in several types of tumors including those found in lung, colon, rectal tumors, and with a particularly high incidence in breast tumors (Grugan et al. 2013; Chiang et al. 2021; Pilley, Rodriguez, and Vousden 2021). This mutation decreases the ability of p53wt to recognize its target genes which are involved in tumor suppression and therefore promotes novel Gain Of Functions (GOF) events through direct or indirect associations with DNA (Grugan et al. 2013; Zawacka-Pankau 2022). This mutation shows some increase in frustration in the NTD and parts of the DBD where ligand binding occurs. Loss of frustration is observed in the CTD, which typically interacts with ligands. The altered frustration in the CTD may reduce p53's ability to interact effectively with DNA and other molecules, leading to diminished tumor-suppressing functions. The increased frustration in the DBD and NTD however apparently enables new binding interactions that contribute to gain-of-function (GOF) properties observed in cancers.

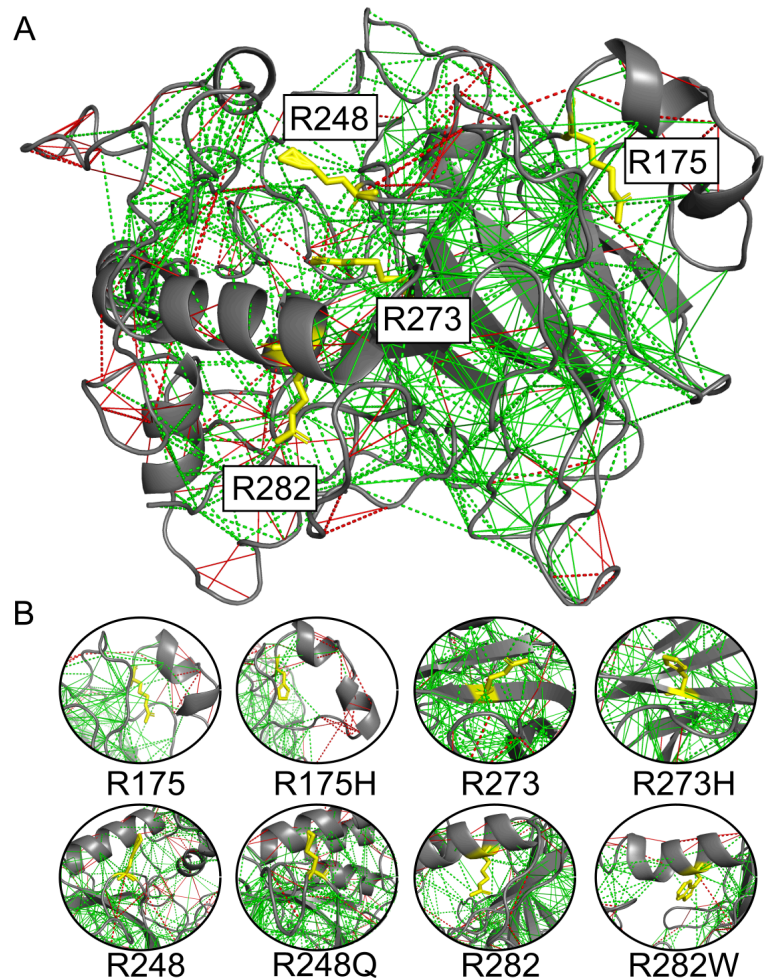

**Figure 14. Cancer causing mutations in p53: A.** Frustration patterns in the p53 native protein, residues corresponding to the discussed mutations are shown in yellow and sticks representation. **B.** Comparisons between frustration patterns between wild-type (left) p53 and each mutation (right). Shown structures are averaged from MD simulations.

The P53R248Q mutation, in which Arg248 is replaced by Gln, is one of the most common mutants found in breast cancers and it is also largely used to predict patient survival (Seagle et al. 2015). This point mutation leads to a weaker binding to DNA with a concomitant increase in aggregation propensity (Yang et al. 2021). The p53R248Q mutation causes the highest increase in frustration among the four studied mutations, particularly in the CTD and near the DNA-binding region. This increased frustration correlates with an increased propensity for the protein to adopt multiple conformations and seems related to an increased aggregation propensity. The flexibility introduced by high frustration also aligns with its involvement in aggressive cancer phenotypes.

The p53R273H mutant, in which Arg273 is replaced by His, is characterized by a reduction in DNA binding capability and it is associated with an inhibition of the autophagy machinery, thereby promoting cancer cell survival (Cai et al. 2015). Similar to p53wt in several regions, p53R273H shows frustration mainly around the DNA-binding domain and some collective motions near the mutation site that exhibit slightly higher frustration than the wild type. The mutated protein also shows some frustration increase at the NTD in the same way that P53R175H does. Overall, the frustration profile does not drastically deviate from p53wt and the mechanism by which frustration may contribute to the phenotypic consequences are not clear. There is some small frustration decrease in the DNA-binding regions which may weaken its interaction with DNA.

Finally, the p53R282W mutant, in which Arg282 is replaced by Trp, is associated with a shorter survival time. This mutant still retains a partial transcriptional ability, however it exhibits a significant aggregation propensity (Yang et al. 2021; Y. Zhang et al. 2016). p53R282W displays high frustration in the DNA-binding domain (particularly near the Zn-coordinated site), where frustration and wide collective motions are observed. This mutation also adds frustration in regions close to the DNA-binding helix. This increased frustration might promote protein aggregation, leading to the formation of fibril-like structures and contributing to shorter patient survival in cancers with this mutation.

### 11.7 Amyotrophic Lateral Sclerosis (ALS)

Amyotrophic lateral sclerosis (ALS) is the most common motor neuron disorder in adults, characterized by progressive degeneration of upper and lower motor neurons, typically resulting in death within five years of symptom onset. While the majority of cases are sporadic, 5%-10% have a familial inheritance pattern, including ALS type 6, which is linked to mutations in the *Fused in Sarcoma* (*FUS*) gene. A study by Bonet and collabs (Bonet et al. 2021) investigated the impact of the most prevalent ALS-related *FUS* mutations—R521C, R521H, and P525L—on protein structure and function. Frustration analysis reveals that the R521C mutation, in particular, leads to increased frustration in the mutated area of the protein. This change disrupts the stability of interactions with surrounding residues, which can impair the protein's normal structure and function. Increased frustration at specific sites may signal a higher likelihood of misfolding or abnormal interaction with chaperones, like protein disulfide-isomerase (PDI), which assists in folding stressed or misfolded proteins. An increase in chaperone binding tendency due to this mutation is further supported by the SNPeffect tool, indicating that the frustration induced by this mutation may induce a greater affinity with PDI. In addition, changes in flexibility and binding affinity reduce the interaction between mutated FUS and its nuclear import receptor, Kapβ2, resulting in the protein's mislocalization to the cytoplasm. This mislocalization is directly linked to ALS's neurodegenerative effects.

### 11.8 Hypertrophic cardiomyopathy (HCM)

Cardiac TnC (cTnC) is a highly conserved protein among mammals that plays a critical role in the calcium-mediated control of muscle contraction, as part of the troponin complex. Genetic variants of this protein can perturb Ca2+-regulation of myocardial contraction and lead to pathological remodeling of the ventricular walls in the form of dilated, hypertrophic, or restrictive cardiomyopathy (DCM, HCM, or RCM, respectively) (Willott et al. 2010; Tadros et al. 2020). cTnC consists of two globular domains, connected by a disordered linker (D/E linker), that binds Ca2+ and/or Mg2+ through EF-hand motifs composed of helix–loop–helix structures.

The N domain binds Ca2+ with lower affinity than the C domain and cTnC samples a range of discrete intermediate states, i.e., the Ca2+-free state (apo) or closed conformation, the Ca2+-bound state or primed conformation, and the Ca2+- and swTnI-bound state or the open, active state. The open state in cTnC is induced when both Ca2+ and the switch TnI segment (swTnI, residues 147–163 in cTnI) bind to N-TnC (Takeda et al. 2003; Li, Spyracopoulos, and Sykes 1999).

A reported variant in hypertrophic cardiomyopathy is C84Y, located at the beginning of the disordered linker between the N and C domains which was reported to increase myofilament Ca2+ sensitivity (Landstrom et al. 2008). Marques and collabs (Marques et al. 2021) reported that the C84Y mutation alters the energy landscape of cTnC by introducing a larger, hydrophobic tyrosine residue at position 84 within the aD-helix. This change disrupts the minimally frustrated network and induces local "high frustration," particularly in the N-domain and the D/E linker regions, which are critical for Ca²⁺ binding and the protein's conformational flexibility. This mutation impacts conformational dynamics of the D/E linker and sampling of discrete states in the N-domain, favoring the "primed" state associated with Ca2+ binding. The authors showed that the cTnC's αD-helix normally functions as a central hub that controls minimally frustrated interactions, maintaining evolutionarily conserved rigidity of the N-domain. αD-helix perturbation remotely alters conformational dynamics of the N-domain, compromising its structural rigidity.

**11.9 Diabetes Type 1 Vaccine:**

A recent article, (Song et al. 2023) studied the effects of altering specific regions of a type 1 diabetes (T1D) autoantigen, the X-idiotype a 16 residues long peptide that is known to activate CD4+ T cells more effectively than insulin, could aid in designing a vaccine to treat this disease. First, the authors performed extensive Molecular Simulations (MD) to obtain stable binary complexes between the X-idiotype and Human leukocyte antigen (HLA) as well as generated tertiary HLA-antigen-TCR complexes by homology modeling. Subsequently, they applied Free Energy Perturbation (FEP) calculations to identify mutations that would increase the binding affinity of the X-idiotype to the HLA. The binding of some X-idiotype residues to HLA were found to be improvable. In particular, 12/13 of the essayed mutations on Tyr6 improved binding, which is compatible with the notion of local frustration. Based on the FEP derived DDGs Tyr6 has a Frustration Index (FI) =1.5. This frustration analysis shows that Tyr6 shows the highest versatility to offer various mutagenesis choices to improve HLA binding and to advance the design of a vaccine to treat T1D.

**11.10 A new inactive conformation of SARS-CoV-2 main protease Mpro: Antiviral Target**

One of the most promising strategies to develop antiviral drug candidates is related to generating protease inhibitors in the form of small molecules that are able to inhibit enzymes involved in virus replication within the cell. Because of the low sequence identity of the viral proteases with their human counterparts as well as their distinct cleavage-site specificities, viral enzymes can be inhibited with very low associated toxic effects ('off-target' effects). The Mpro protein is the main protease in coronaviruses. It is a cysteine peptidase that is essential in positive sense, single-stranded RNA coronaviruses that need it for their replication cycle (Xia and Kang 2011). In a recent study, a new inactive conformation of the SARSCov2 MPro protein was described as a weak point of the virus that may enable the development of antivirals against it (Fornasier et al. 2022). The authors processed different data sets to determine crystal structures of Mpro in complexes with different inhibitors. After processing the data and solving the structures they found that, in some cases, some portions of the protein were difficult to model as they were in a different conformation from the ones known in other experimental structures. They described this new conformation as a newly identified "inactive" state (termed "new-inactive"), where critical regions such as the oxyanion loop shift into a catalytically incompetent configuration. By analysing the frustration patterns in the active and new-inactive conformations of MPro, the authors observed that Cys145 establishes 18 minimally frustrated interactions in the new-inactive conformation vs 8 in the active one. Cys117 seems to play an important role in the stabilization of the new-inactive conformation. Within the oxyanion loop there is highly frustrated interaction between Leu141 and Ser139 in the new-inactive conformation while in the active conformation Leu141 is highly frustrated with respect to Ser144. This suggests that Leu141 is important in switching between the two conformations. In Addition, Phe140 shows no minimally frustrated interactions in the new-inactive conformation and eight in the active one, suggesting that it is needed to stabilize the active conformation. Based on this observation, it could be possible to design inhibitors that could lock Mpro in its inactive state Targeting Phe140 or Leu141 to prevent the enzyme from reverting to its active conformation.

## 11.11 Large Scale Analysis of Frustration in Naturally Occurring Human Mutations

All these examples put into context how local frustration can change upon mutations occurring in proteins which can lead to the diversification of their biophysical properties not only over the longtime of evolution but also more acutely give rise to pathogenic behaviors. An obvious objective would be to dissect general rules within protein mutations that would allow us to classify them as benign or pathogenic, leading to better ways to screen the potential effect of such variants when screened in human samples, for example.

Oncogenic mutations in protein kinases can alter the distribution of locally frustrated sites within the kinase structure. These locally frustrated regions are areas where interactions are not energetically optimized and are thus prone to dynamic modulation. Dixit et. al. found that cancer mutations may increase local frustration in the inactive state of kinases while reducing frustration in the active state (Dixit and Verkhivker 2011). This shift in frustration patterns can promote allosteric transitions between functional states, potentially leading to constitutive kinase activation - a hallmark of many cancers. Locally frustrated sites may serve as catalysts for conformational changes induced by cancer mutations, facilitating the shift towards an active kinase state. This interplay between oncogenic mutations and local frustration provides insight into how cancer-associated changes in protein kinases can alter their energy landscapes and activation mechanisms.

A more recent study also analyzed local frustration patterns associated with mutations and their impact on the biophysical properties of naturally occurring human proteins, ultimately leading to pathogenic phenotypes. A comprehensive analysis of sequence variations in frustration patterns was conducted using a large dataset of rare single-nucleotide variants in human genome coding regions (Kumar, Clarke, and Gerstein 2016). The study revealed that disease-related single-nucleotide variants cause more substantial changes in localized frustration compared to non-disease related variants. Additionally, rare single-nucleotide variants were found to disrupt local interactions more significantly than common variants. Notably, somatic single-nucleotide variants linked to oncogenes and tumor suppressor genes exhibited distinct alterations in frustration patterns. Mutations in tumor suppressor genes induced greater frustration changes in the core rather than the surface, resulting in loss-of-function events. Conversely, oncogene-associated mutations displayed the opposite effect, creating gain-of-function events, sometimes by alleviating frustration and interfering with proper allosteric communication (Kumar, Clarke, and Gerstein 2016).

## 12. Molecular Frustration as a Regulatory Principle in Viral Infections

Viruses are complex systems that self-assemble at the end of their viral life cycle. This is a directional process that starts with the infection of a cell and ends with the release of multiple progeny virions that can repeat the process in other susceptible cells. After a virion enters a cell, it becomes sensitive to different stimuli from the host environment that results in sequential release of its genome, subsequent gene expression and the formation of new virions. Recent work (Twarock, Towers, and Stockley 2024) proposes that molecular frustration is part of a fundamental mechanism that directs the different sequential steps from genome reseal to viral assembly.

In the case of the MS2 bacteriophage, a single stranded (ss)RNA virus, the viral genome (gRNA) contains multiple packaging signals (PSs) that interact with coat proteins (CPs) to direct capsid formation. However, after the assembly is complete, many of these PS–CP interactions are selectively disrupted, generating local strain within the capsid. This "controlled" instability ensures that the virion remains stable enough to protect the genome while remaining primed for efficient uncoating upon host cell entry. The MS2 coat dimer adopts two conformations (A/B and C/C) that coexist and are present in a ratio of 2:1 when the capsid is fully assembled (Fig. 15A). There is a specific region that undergoes a conformational change in the monomers of the coat protein, the FG-loop. In the B unit, a proline in position 78 (Pro78) adopts a cis conformation making the FG-loop bend towards the globular body of the A/B dimer, allowing it to fit the capsid without clashes. In the A or C units, Pro78 adopts a trans conformation that confines the FG-loop to be in the A and C conformation. In solution the C/C conformation dominates while when it interacts with the gRNA PSs the conformational equilibrium is shifted towards the A/B form (Fig. 15B). A frustration comparison between the coat protein both in B and A subunits (Fig. 15C) shows that there is significantly less frustration in the B subunit in comparison with the A subunit. This supports the notion that the PS interactions with the coat protein results in one of the C subunits adopting the B conformation but also in a decrease of frustration in the other monomer of the CP dimer via allosteric regulation. Finally, the authors of the study hypothesize that the assembled version of the virion, composed of capsid and packaged gRNA, can be understood as an ensemble or alternative states, characterized by different numbers of interactions between the CP and the PS-gRNA as well as distinct conformations of the packaged gRNA. The interplay between bound and unbound conformations between the gRNA and the CP molecules allow the tertiary structure of the gRNA to be reorganized in a more favourable configuration. Therefore, analogously to the folding process, the virion ensemble of alternative states follows the minimum frustration principle which results in a coordinated and programmed breaking of PS-CP contacts that gives as a result a successful infection at the correct time and location.

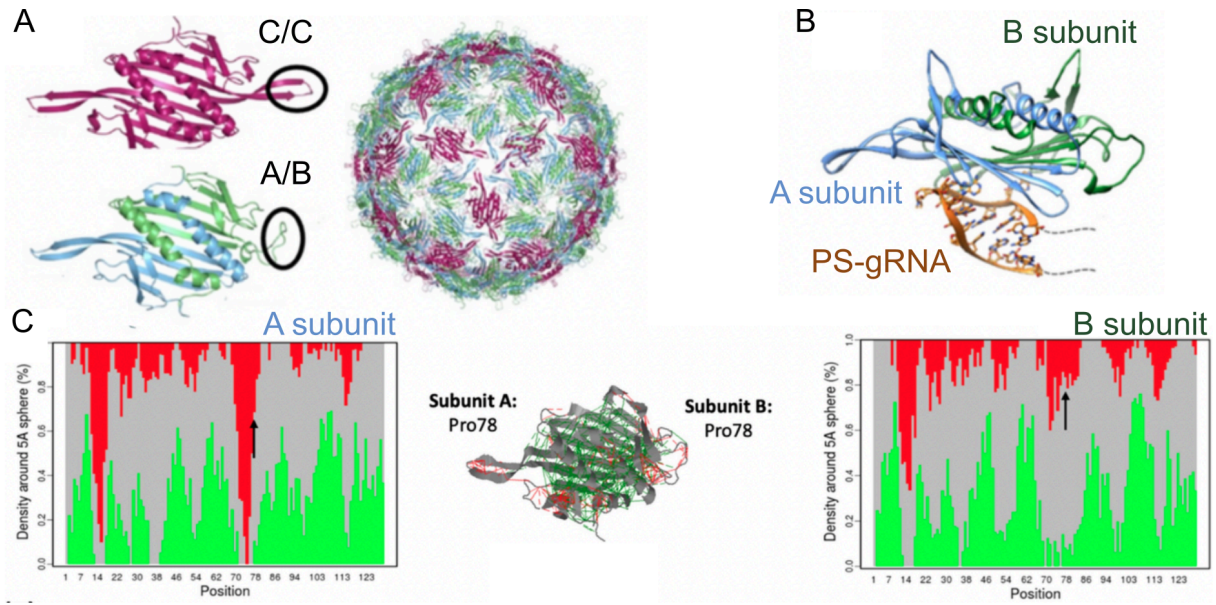

**Figure 15: PS-gRNA interactions with the CP in MS2 leads to conformational and frustration changes.** These frustration changes result in the virion disassembly and release of the gRNA genome in the cell. **A.** Schematic view of the C/C and A/B dimer forms of the MS2 CP, **B.** Complex between the A/B conformation and the PS-gRNA. **C.** Configurational frustration comparison between the A and B subunits in the A/B dimer showing differential frustration, especially around Pro78. Figure modified from (Twarock, Towers, and Stockley 2024) licensed under the Creative Commons CC-BY international license (http://creativecommons.org/licenses/by/4.0/).

Similar PS-mediated assembly mechanisms have been observed in other single-stranded ssRNA, including picornaviruses and alphaviruses, where interactions between the CPs and the genome define a preferred assembly pathway that prevents premature disassembly. Additionally, a related form of frustration by incorporating specific RNA signals that regulate nucleocapsid maturation, has been proposed in the case of the hepatitis B virus (HBV) demonstrates, ensuring that the virus remains stable until it reaches the cytoplasm. This dual role of molecular frustration, facilitating both precise assembly and programmed disassembly, suggests that it plays a broader role in regulating viral life cycles.

In the case of the HIV virus, its capsid must maintain its integrity as it traffics through the cytoplasm and across the nuclear pore, yet it must also disassemble at the appropriate moment to release the viral genome. The virus achieves this balance through interactions with host cofactors, such as Cyclophilin A and FG-nucleoporins, which modulate capsid stability by reducing molecular frustration at key structural sites. This stepwise reduction of frustration provides a mechanism for infection directionality, ensuring that uncoating occurs only when the virus reaches its replication site. This concept challenges the traditional view of viral capsids as static containers, instead highlighting their dynamic, energy-dependent regulation.

The implications of this novel role of molecular frustration extend beyond mechanistic insights about viral assembly to potential therapeutic applications. By targeting PS–CP interactions or modulating host cofactor binding, it may be possible to design novel antiviral strategies that interfere with frustration-mediated viral regulation. Small molecules that either stabilize or destabilize capsid structures inappropriately could effectively hinder viral infectivity. Additionally, insights from frustration-driven assembly mechanisms could improve the design of viral vectors for gene therapy, ensuring more efficient genome packaging and controlled uncoating. This concept may prove to be a universal framework for understanding and manipulating viral infections at a molecular level.

## 13. FRUSTRATION IN GENE REGULATORY NETWORKS

The energetic analysis of protein structural dynamics is underpinned by the fact that once assembled proteins are never very far from equilibrium. The attractors of such near equilibrium systems are static and can be described by an energy landscape. Brains are not entirely at equilibrium, so there are many patterns of attractors, not just stable firing patterns. These are manifested as oscillations such as their various rhythms detected by an EEG, amongst others. Where do gene networks lie in this spectrum of behavior of stasis versus oscillation? Certainly, there are some gene networks that oscillate when they function. The NFκB-IκBα gene circuit is an important example in immunology. Stimulating the NFκB/IκBα system, as happens in response to inflammation, leads to an oscillation with a period of less than a day, the "ultradian" rhythm. The ultradian oscillatory response of this network correlates with many clinical features of fighting infections. The response is started by degrading the inhibitor IκBα, unleashing the NFκB to which it has been bound in the cytoplasm. The now freed NFκB is a transcription factor that turns on many genes to fight the infection, one of which however is IκBα itself. The synthesis of IκBα then starts to turn off the transcriptional activity NFκB by stripping it from its many genetic sites. Failure to turn off the response could lead to autoimmune diseases. We see that this genetic regulatory loop is an example of a frustrated interaction between NFκB and IκBα expression, leading to the oscillation. Doubtless, there are more sub-networks involved in gene regulation that are frustrated in a similar way.

The structure and dynamics of gene regulatory networks is only beginning to be investigated through the lens of energy landscapes and frustration analysis. Sasai and Wolynes looked into the question of the counting of cell types (Sasai and Wolynes 2003). Cell types are often caricatured as mere patterns of gene expression. A large number of genes are involved in setting up a cell type, and it is possible to imagine exponentially many patterns of the interactions among the genes, either repressing each other or stimulating each other, by communicating through their transcription factors. If there are N genes, then one might envision there would be $2^N$ possible gene states. The number of cell types, while large, does not appear to be so exponentially large. Sasai and Wolynes explained this relative paucity of cell types by building a stochastic model gene network that could be mapped onto the Hopfield magnet model of the neural networks. Like the Hopfield magnet, when the frustration level of the network is relatively low, a small number of stable (static) cell types emerges with stable expression patterns being only a fraction of the number of genes. They suggested minimal frustration then was a key aspect of the gene networks that are involved in cell type selection.

Recently, using expression data Tripath, Kessler and Levine have confirmed this idea (Shubham Tripathi, Kessler, and Levine 2020). They analyzed genetic regulatory networks connected with several genetic networks including the messenchymal-epithelial transition which is so important in cancer metastasis. They also looked at the pluripotency network and the MYC network. They have concluded the pathways of these networks are all minimally frustrated in the sense of the Sasai-Wolynes model. They emphasize that minimally frustrated gene networks are robust to random variations of their parameters and presumably remain stable under varying environmental influences making them an essential aspect of physiology.

The frustration concept would appear to provide a new tool then giving a new way of studying regulatory diseases and may allow us to pinpoint specific genes where dysfunction in particular networks can be repaired.

## 14. FRUSTRATION IN MOLECULAR PSYCHIATRY

In this review, we have explained how the seemingly psychological concept of "frustration" can be translated into quantitative terms for physical systems, such as glasses, biomolecules and their assemblies. In the latter systems their functions hinge upon long lived but near equilibrium states so the concept of an energy landscape with a statistical character can be used and translated into mathematics. This raises the question "can the mathematical analysis of frustration be taken back to brain physiology and, perhaps, be useful to psychiatry?"

Some work in this direction has emerged but clearly there is much that needs to be clarified. As we have seen, in the introduction concerning artificial neural networks, the long lived states that are needed to describe long term biological memory and that underlie the ability of brains to generalize can be imitated by near equilibrium systems for which the energy landscape concepts can directly be applied. Yet, other essential brain activities require more organized action and rapid dynamical responses in real time. Thus, oscillatory and chaotic behaviors are manifested in the psychological brain circuitry of humans. While the quantitative notion of frustration can be taken over for oscillatory behavior, as we have seen for the NFκB/IκB oscillatory gene circuit, it is not yet established how best to quantify frustration for complex many body systems that do not possess quasi-stable states. The necessity of introducing non gradient flows described by gauge fields has been spotlighted by Wang, as being critical to model systems that are very far from equilibrium (J. Wang 2015).

Friston has nevertheless argued that a very significant part of normal brain function is in fact governed by a free energy principle. This principle is based on the idea that Bayesiam inference is critical for action, perception and learning (Friston 2010, Friston 2023). Going further, Tozzi et al. have argued that the brain, being described by a free energy principle, is minimally frustrated (Tozzi, Flå, and Peters 2016). They have usefully taken over parts of the energy landscape theory of protein folding to discuss information flow in the brain.

Gollo and Breakspear have suggested that the local tuning of the frustration levels in brain circuits is critical to neuronal integration (Gollo and Breakspear 2014). They point to local patterns of triplets of neural centers which can either be frustrated in their ability to synchronize or cooperate to synchronize. Epilepsy is manifested in unusual patterns of over synchronization (Stafstrom and Carmant 2015) during a seizure. A seizure according to this picture then comes from there being too little frustration in the circuits leading to a phase transition to the synchronized state. Carhart-Harris and his collaborators have argued that psychedelic drugs, by acting as agonists of serotonin receptors, re-tune the balance between inhibition and excitation in neuronal networks, resulting in more chaotic behavior (Carhart-Harris and Friston 2019). When such drugs are used such a change from simple recall to "tripping" then would resemble what happens to the Hopfield associative memory network when it is overtrained, leading to frustrated interactions that, in turn, cause hallucinations.

## 15. SUMMARY

We hope to have shown how the concept of frustration can give insight in physiology and molecular medicine. At the biomolecular level, a statistical interpretation of energetic frustration has made possible a way of quantifying the selection value of tuning frustration. Minimal frustration is encoded into most protein sequences so that the proteins can organize three dimensionally into stable structures readily. At the same time, evolution has introduced frustration locally into protein molecules so that once they are folded, they can bind specific partners and undergo functional motions. Easy to use algorithms for quantifying frustration have been developed and have been applied across the biomolecular world. Very often, mutations causing changes of frustration patterns are signals of the sites of pathology. Frustration analysis may also provide clues to finding druggable sites (M. Chen et al. 2020) and candidate drugs. Wang has argued that the most specific drugs have strongly funneled binding landscapes (Yan and Wang 2020) and form minimally frustrated complexes in contrast to drugs binding to many targets.

The frustration concept is also being explored at higher levels of biological organization (Wolf, Katsnelson, and Koonin 2018). At the cellular level it appears that many gene regulatory networks have evolved to be minimally frustrated (Shubham Tripathi, Kessler, and Levine 2020). The ideas of frustration and energy landscapes are also starting to re-enter brain research and psychiatry. It will be exciting to see how understanding these higher order aspects of frustration will develop and contribute medicine in coming years.

### CRediT authorship contribution statement

R. Gonzalo Parra: Conceptualization, Visualization, Writing – original draft, Writing – review & editing. Elizabeth A. Komives: Conceptualization, Funding acquisition, Writing – original draft, Writing – review & editing. Peter G. Wolynes: Conceptualization, Funding acquisition, Writing – original draft, Writing – review & editing. Diego U. Ferreiro: Conceptualization, Funding acquisition, Writing – original draft, Writing – review & editing.

**Data availability statement**

The original contributions presented in the study are included in the article. Further inquiries can be directed to the corresponding authors.

**Funding**

RGP was supported by a fellowship from Grant IHMC22/00007/PRTR funded by the Instituto de Salud Carlos III (ISCIII). DUF was supported by the Consejo de Investigaciones Científicas y Técnicas (CONICET); CONICET Grant PIP2022-2024—11220210100704CO and Universidad de Buenos Aires grant UBACyT 20020220200106BA. PGW was supported both by the Bullard–Welch Chair at Rice University, grant C-0016, and by the Center for Theoretical Biological Physics sponsored by NSF grant PHY-2019745. EAK was supported by NIH grant R01HL127041.

**Conflict of interest**

The authors declare that there are no commercial or financial relationships that could be construed as a potential conflict of interest.